\documentclass{article}

\usepackage{arxiv}
\usepackage[utf8]{inputenc} 
\usepackage[T1]{fontenc}    
\usepackage{hyperref}       
\usepackage{url}            
\usepackage{graphicx} 
\usepackage{subcaption}
\usepackage{amssymb}
\usepackage{multirow}
\usepackage{float}
\usepackage{amsmath}
\usepackage[table]{xcolor}
\usepackage{threeparttable}
\usepackage{parskip}
\usepackage{soul}
\usepackage{pdflscape}
\usepackage{pifont} 

\title{A Survey of Learning-Based Intrusion Detection Systems for In-Vehicle Network}

\author{ \href{https://orcid.org/0000-0002-2150-7981}{\includegraphics[scale=0.06]{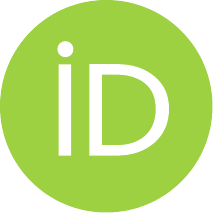}\hspace{1mm}Muzun Althunayyan} \\
	School of Computer Science \& Informatics\\
	Cardiff University\\
	Cardiff, United Kingdom \\
	\texttt{AlthunayyanMS@cardiff.ac.uk} \\
	\And
	\href{https://orcid.org/0000-0001-9761-0945}{\includegraphics[scale=0.06]{orcid.pdf}\hspace{1mm}Amir Javed} \\
	School of Computer Science \& Informatics\\
	Cardiff University\\
	Cardiff, United Kingdom \\
	\texttt{javeda7@cardiff.ac.uk} \\
 \And
	\href{https://orcid.org/0000-0003-3597-2646}{\includegraphics[scale=0.06]{orcid.pdf}\hspace{1mm}Omer Rana} \\
	School of Computer Science \& Informatics\\
	Cardiff University\\
	Cardiff, United Kingdom \\
	\texttt{ranaof@cardiff.ac.uk} 
}

\begin{document}
\maketitle

\begin{abstract}
Connected and Autonomous Vehicles (CAVs) enhance mobility but face cybersecurity threats, particularly through the insecure Controller Area Network (CAN) bus. Cyberattacks can have devastating consequences in connected vehicles, including the loss of control over critical systems, necessitating robust security solutions. In-vehicle Intrusion Detection Systems (IDSs) offer a promising approach by detecting malicious activities in real time. This survey provides a comprehensive review of state-of-the-art research on learning-based in-vehicle IDSs, focusing on Machine Learning (ML), Deep Learning (DL), and Federated Learning (FL) approaches. Based on the reviewed studies, we critically examine existing IDS approaches, categorising them by the types of attacks they detect—known, unknown, and combined known-unknown attacks—while identifying their limitations. We also review the evaluation metrics used in research, emphasising the need to consider multiple criteria to meet the requirements of safety-critical systems. Additionally, we analyse FL-based IDSs and highlight their limitations. By doing so, this survey helps identify effective security measures, address existing limitations, and guide future research toward more resilient and adaptive protection mechanisms, ensuring the safety and reliability of CAVs.
\end{abstract}


\section{Introduction}
Connected and Autonomous Vehicles (CAVs) are expected to become the backbone of future transportation systems \cite{aloraini2024adversarial}, offering the potential to not only revolutionise mobility, but also deliver significant economic benefits. For example, the Society of Motor Manufacturers and Traders (SMMT) estimates that this technological shift could provide the United Kingdom with an annual economic boost of £62 billion by 2030 \cite{SMMT2019Report}. However, these advancements also present significant security challenges for CAVs, making them attractive targets for emerging cyber threats \cite{pickford2024systematic}.

CAVs rely on Electronic Control Units (ECUs) to manage and control various functions. These ECUs communicate through standardised in-vehicle communication protocols, such as the Controller Area Network (CAN), FlexRay, Local Interconnect Network (LIN) and Media Oriented System Transport (MOST). Among these, the CAN bus is the protocol that is the most widely used, valued for its high speed, reliability, and ease of use \cite{al2019intrusion}. Although originally designed for industrial applications, the CAN bus has become the de facto standard for in-vehicle communication \cite{althunayyan2024robust}. Despite its advantages, the CAN protocol was not designed with security in mind and lacks essential features such as sender authentication and encryption \cite{paul2021artificial}. 

The increasing interconnectivity of CAVs exposes them to a range of cyberattacks. The attack surfaces in modern vehicles can be accessed either physically, via ports such as the USB or the onboard diagnostic (OBD)-II port, or remotely through wireless technologies such as Bluetooth, Wi-Fi, and LTE \cite{aliwa2021cyberattacks}. In 2023, the number of large-scale incidents, potentially affecting thousands to millions of mobility assets, grew 2.5-fold compared to 2022. Additionally, 95\% of cyberattacks are conducted remotely, with 85\% being long-range \cite{Upstream2025report}. These vulnerabilities make vehicles susceptible to attacks that could have devastating consequences, including loss of control over critical systems like braking, steering, and acceleration \cite{young2019survey}. A recent incident \cite{Ken2023keylesstheft} involved cybersecurity researcher Ian Tabor discovering tampering on his Toyota RAV4, particularly around the front bumper and headlight area. Initially suspecting vandalism, Tabor soon realised the vehicle had been targeted by a cyberattack. Investigations revealed that attackers had accessed the car's CAN bus through exposed wiring, allowing them to inject malicious signals. This manipulation enabled the attackers to unlock the doors and start the engine, ultimately stealing the vehicle without the need for a key. Moreover, a notorious example is the Jeep hack, where attackers remotely gained control over the vehicle’s braking and steering systems, resulting in dangerous driving conditions \cite{golson2016jeep}. Similarly, vulnerabilities in BMW and Toyota Lexus models have been exploited, demonstrating the persistent threat to vehicle security \cite{tencentexperimentalBMW, tencentexperimental}. Such incidents underscore the need for robust security measures to protect against both information theft and direct physical harm.

Given the severity of these threats, the security of the CAN bus has become a major area of research. According to McKinsey’s analysis, by 2030, almost 95\% of the new vehicles will be connected to external networks, further highlighting the need for effective security solutions \cite{bertoncello2021unlocking}. One promising approach is the implementation of \textbf{Intrusion Detection Systems (IDSs)}, which monitor network traffic for malicious activity. In the context of in-vehicle networks, an IDS is typically installed on an ECU and analyses incoming messages to detect abnormalities. However, conventional IDS technologies designed for traditional networks cannot be applied directly to in-vehicle systems due to resource constraints and the real-time requirements of automotive environments.

Research on the development of in-vehicle IDSs has expanded considerably in recent years, fueled by the discovery of various vulnerabilities and the urgent need to improve the security of in-vehicle networks and detect cyberattacks. Researchers have explored various approaches to building these systems. IDSs can be classified as either signature-based, for detecting known attacks, or anomaly-based, for identifying new, unknown attacks ~\cite{hoppe2009applying}. Anomaly-based IDSs are further categorised into statistical, Machine Learning (ML), rule-based, and physical fingerprinting methods ~\cite{rajapaksha2023ai}.

This survey specifically examines the development of learning-based in-vehicle IDSs, with a focus on Machine Learning (ML), Deep Learning (DL), and Federated Learning (FL) approaches. It aims to identify effective security strategies, overcome existing challenges, and guide future research toward more robust and adaptive protection mechanisms to ensure the safety and reliability of CAVs. The emphasis on ML- and DL-based approaches is driven by their strong generalization capabilities and ability to process large volumes of traffic data ~\cite{rajapaksha2023ai}. Additionally, FL has recently gained attention among researchers due to its potential to enhance both security and privacy.

In this survey, we review the state-of-art research on ML-based, DL-based and FL-based in-vehicle IDSs, aiming to identify limitations and research gaps in the existing work.

\textbf{Contribution} 
 To summarise, the contributions of this paper are as follows:

\begin{itemize}
    \item We present a comprehensive literature review employing a structured search strategy to systematically gather research papers published up to January 2025.

     \item We present a systematic categorization of CAN protocol vulnerabilities by conducting an in-depth analysis of the CAN protocol, identifying its weaknesses, entry points, and potential attack scenarios.

    \item We introduce a classification framework for IDS methodologies based on the types of attacks they detect, including known, unknown, and combined known-unknown threats. Additionally, we present summary tables and highlight the limitations of each approach to identify research gaps.

    \item We analyse the evaluation metrics used in research studies, emphasising the importance of considering multiple factors—such as performance, time complexity, and memory overhead—when developing in-vehicle IDSs to ensure they meet the requirements of safety-critical systems.

   \item We provide insights into FL-based IDSs, discussing their advantages while also identifying limitations in addressing security and privacy challenges in connected vehicle environments.

   \item We outline key open challenges and propose future research directions to enhance the development of more effective, efficient, and resilient in-vehicle IDS solutions.

\end{itemize}
The remainder of this paper is organised as follows. Section~\ref{Background} provides context and background information. Section~\ref{Related_Work} reviews similar surveys and highlights our contributions. Section~\ref{Methodology} outlines the search methodology used to collect relevant papers. Section~\ref{Existing_IDSs} reviews existing ML- and DL-based in-vehicle IDSs, while Section~\ref{Federated_Learning_Section} focuses on FL-based IDSs. Section~\ref{Future_Research_Directions} discusses future research directions. Finally, Section~\ref{Conclusion} concludes the paper.

\section{Background}
\label{Background}
This section provides a brief overview of in-vehicle protocols, with a particular focus on the CAN protocol, including its description, functionality, and aspects relevant to cyberattacks. In addition, it examines the vulnerabilities, entry points, and attack scenarios of CAN.

\subsection{In-vehicle Network}
\label{In-vehicle_Network}
The in-vehicle network is the internal network that facilitates communication between multiple ECUs within a vehicle \cite{liu2020client}. These ECUs are interconnected embedded devices responsible for controlling various vehicle functions, such as engine management, airbag control, and climate control. The number and type of ECUs in a vehicle vary depending on the manufacturer and model, with modern vehicles incorporating up to 100 ECUs alongside basic functions \cite{ahmad2024comprehensive}. These ECUs, along with sensors, actuators, cameras, radars, and communication devices, collaborate to enhance vehicle performance, efficiency, intelligent services, and safety by collecting and interpreting various data~\cite{boumiza2017intrusion}. ECUs communicate using standard protocols such as CAN, FlexRay, LIN, and MOST~\cite{kumar2014automotive}. Among these, CAN is considered the de facto protocol for in-vehicle communication~\cite{al2019intrusion}.

\subsection{Controller Area Network}
\label{CAN_protocol}
The Controller Area Network (CAN) protocol, developed by Robert Bosch in 1985, was designed to reduce the weight, complexity, and cost of wiring. Due to its high speed and efficiency, CAN has become the most widely used in-vehicle communication protocol in connected and autonomous vehicles~\cite{al2019intrusion}. CAN operates as a message-based broadcast protocol, where the ECUs transmit data in pre-defined frames. Since the system uses a broadcast mechanism, each message is sent to all ECUs on the network.

\subsection{CAN Bus Data Frame}
\label{CAN_Bus_Data_Frame}
\begin{figure}[!ht]
\centering
    \includegraphics[scale= 0.18]{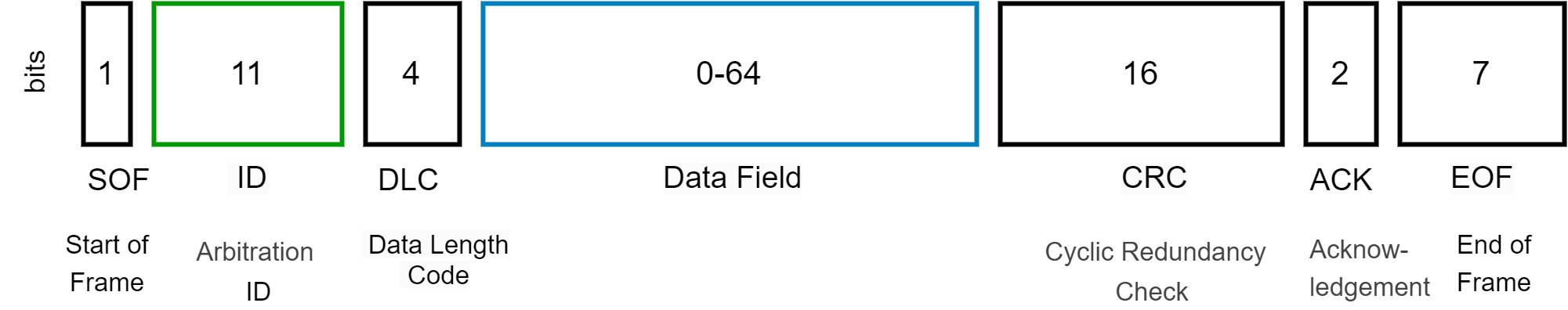}
    \caption{CAN data frame}
    \label{fig:CANDataFrame}
\end{figure}

The CAN frame follows a specific message structure defined in a database-like file known as the DataBase CAN (DBC) file. This file is confidential and proprietary to the vehicle manufacturer, containing all the essential information about the representation of the CAN bus data~\cite{lokman2019intrusion}. The CAN data frame consists of seven fields that facilitate data transmission between ECUs. Figure ~\ref{fig:CANDataFrame} illustrates the standard CAN frame format, which consists of the following fields:

\begin{itemize}
  \item \textbf{Start of Frame (SOF):} The purpose of this field is to synchronise the transmission of the CAN message with all nodes and to signal the initiation of its transmission. 

  \item \textbf{Arbitration Field (ID):} This field, also known as the CAN ID, is used to specify the destination address of the designated ECU. It also determines the priority of the message, where a lower value generally indicates a higher priority. The ID field is 11 bits in size. 

  \item \textbf{Data Length Code (DLC):} This field provides information about the length of the data field.  

  \item \textbf{Data Field:} This field, also known as the payload, includes the actual vehicle parameter values, which are interpreted by the received ECU and its size can vary from 0 to 8 bytes (0-64 bits).

  \item \textbf{Cyclic Redundancy Check (CRC):} This field detects errors and maintains data integrity during message transmission with a fixed size of 16 bits. 

  \item \textbf{Acknowledge Field (ACK):} This field receives confirmation from the receiving node that the CAN message was received correctly. 

  \item \textbf{End of Frame (EOF):} This field signifies the completion of CAN message transmission.
\end{itemize}

\subsection{CAN Vulnerabilities}
\label{CAN_Vulnerabilities}
The CAN bus was introduced to reduce costs, simplify installation, and improve real-time communication efficiency within vehicles. However, it is vulnerable to cyberattacks due to several inherent vulnerabilities~\cite{aliwa2021cyberattacks, carsten2015vehicle, liu2017vehicle}, including the following:

\begin{itemize}
    \item \textit{Lack of authentication:} Due to the lack of authentication on the CAN bus, any ECU can transmit a frame using the CAN ID of another ECU \cite{young2019survey}. Each ECU broadcasts and receives all data on the bus, then determines whether a message is intended for it. However, the CAN protocol is inherently unable to prevent unauthorised devices from joining the network and sending malicious messages to all ECUs. As a result, attackers can exploit compromised ECUs to spoof and send fake CAN packets, leading to spoofing and message injection attacks \cite{rajapaksha2023ai}.
    
    \item \textit{Lack of encryption:} Due to time constraints, CAN messages are not encrypted \cite{aliwa2021cyberattacks}, allowing cyberattackers to easily capture and analyse them for further attacks. Lack of encryption makes CAN traffic vulnerable to sniffing, spoofing, modification, and replay attacks \cite{young2019survey}.
     
    \item \textit{Broadcast domain:} The CAN bus functions as a broadcast domain, where all ECUs receive the transmitted frames. Each ECU then checks the data and determines whether to process or disregard it \cite{ahmad2024comprehensive}.  If an ECU is compromised, it can intercept and monitor all messages transmitted across the CAN network, enabling an eavesdropping attack \cite{dupont2019survey}.
    
    \item \textit{ID-based priority:} The CAN network prioritises messages based on their IDs, with lower IDs having higher priority \cite{rajapaksha2023ai}. Attackers can exploit this by repeatedly sending frames with low IDs, resulting in a Denial-of-Service (DoS) attack \cite{lokman2019intrusion}.

    \item \textit{Unsegmented Network:} All ECUs are connected to a single shared network without segmentation \cite{young2019survey}, which is a key reason why CAN was adopted in automotive systems, as it eliminates the need for point-to-point connections. However, this shared network allows components such as infotainment systems to communicate with safety-critical vehicle systems. Although some manufacturers use separate networks for safety-critical systems, there is still cross-communication between critical and non-critical systems \cite{young2019survey}.

    \item \textit{External Interfaces:} The attack surface of the CAN bus network is expanded by external interfaces such as the OBD-II port, used for vehicle maintenance and diagnostics; the Telematics Unit, which provides connectivity to the vehicle via Wi-Fi, Bluetooth, GPS, and mobile data interfaces; and the Infotainment Unit, which delivers information and entertainment to the driver through a head display unit, including features like CD/DVD players and USB ports. These interfaces create additional entry points for potential cyberattacks~\cite{wu2019survey, rajapaksha2023ai}.

\end{itemize}

\subsection{In-vehicle Network Entry Points}
\label{entry_points}
Attackers can gain access to the CAN bus or specific ECUs either physically or remotely\cite{aliwa2021cyberattacks, limbasiya2022systematic, checkoway2011comprehensive, carsten2015vehicle}. These entry points serve as gateways for initiating a range of attacks, exploiting the inherent vulnerabilities described in Section \ref{CAN_Vulnerabilities} in in-vehicle networks. This section discusses the entry points attackers can exploit to access the in-vehicle network, either physically or remotely.

\subsubsection{Physical Access} 
Physical access allows an attacker—such as a mechanic, valet, car renter, or anyone with even brief access to the vehicle—to directly interact with its internal systems. This access, even for a short time, can provide opportunities to exploit vulnerabilities through various physical entry points, including:

\begin{itemize}
   \item \textbf{OBD-II Port:} The OBD-II port, commonly located under the dashboard in most vehicles, provides the simplest and most direct access to a vehicle’s primary CAN buses. This port offers sufficient access to potentially compromise the full range of automotive systems \cite{checkoway2011comprehensive}. Designed primarily for vehicle maintenance and engine diagnostics, the OBD-II port allows mechanics to connect scanning tools and capture data packets generated by malfunctioning subsystems. Despite its intended purpose, the port’s accessibility makes it a significant security vulnerability. Attackers can easily connect to the OBD-II port to extract information or install malware onto the vehicle’s systems, disconnecting afterward to leave no physical evidence \cite{koscher2010experimental}. Alternatively, attackers may deploy a remote device to the port, or enabling continuous data collection or exploitation over time. Since the OBD-II port is required for maintenance and diagnostics, it will always pose a security risk \cite{carsten2015vehicle}.

   \item \textbf{Aftermarket Components:} Peripheral components such as USB ports, CD players, and third-party add-ons also pose security risks \cite{liu2017vehicle, carsten2015vehicle}. For example, malicious devices, including FM radios, USB connectors, or CD players purchased from unverified or aftermarket sources, can introduce malware into the vehicle’s system. While these components may be more affordable, they can compromise the vehicle’s security \cite{koscher2010experimental}.
\end{itemize}

\subsubsection{Remote Access}
An external attacker can exploit wireless interfaces commonly implemented in modern vehicles, such as Bluetooth, Wi-Fi, cellular networks, and GPS, without requiring physical proximity to the vehicle. Once these interfaces are accessed, the attacker can inject malicious data or commands into the CAN bus \cite{aliwa2021cyberattacks}. Koscher et al. \cite{chockalingam2016detecting} highlight the feasibility of executing various types of wireless attack injections on in-vehicle network systems. For example, vulnerabilities in telematics systems or vehicle-to-cloud communications can enable the remote injection of messages, disrupting the network. Specific methods include using malicious Windows Media Audio (WMA) files or sending malicious packets to the telematics unit via 3G Internet Relay Chat (IRC) \cite{chockalingam2016detecting}. Moreover, Woo et al. \cite{woo2014practical} conducted a wireless attack, successfully taking control of a target vehicle by utilising malware installed on a smartphone. These examples highlight the significant security risks posed by wireless interfaces in connected vehicles.

\subsection{Attack Scenarios}
\label{Attacks_Scenario}
Since an attacker can gain access to the in-vehicle network, either physically or remotely, through one of the entry points outlined in Section~\ref{entry_points}, the following are common attack scenarios:

\begin{figure}[t]
\centering
\begin{subfigure}{0.28\textwidth}
    \includegraphics[width=\textwidth]{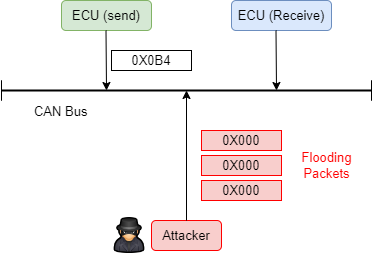}
    \caption{DoS attack}
    \label{fig:flooding}
\end{subfigure}
\hfill
\begin{subfigure}{0.28\textwidth}
    \includegraphics[width=\textwidth]{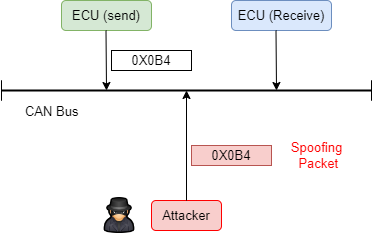}
    \caption{Spoofing attack}
    \label{fig:spoofing}
\end{subfigure}
\hfill
\begin{subfigure}{0.28\textwidth}
    \includegraphics[width=\textwidth]{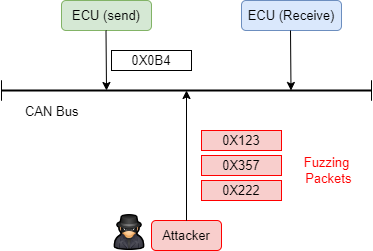}
    \caption{Frame fuzzification attack}
    \label{fig:fuzzuing}
\end{subfigure}
\hfill
\begin{subfigure}{0.28\textwidth}
    \includegraphics[width=\textwidth]{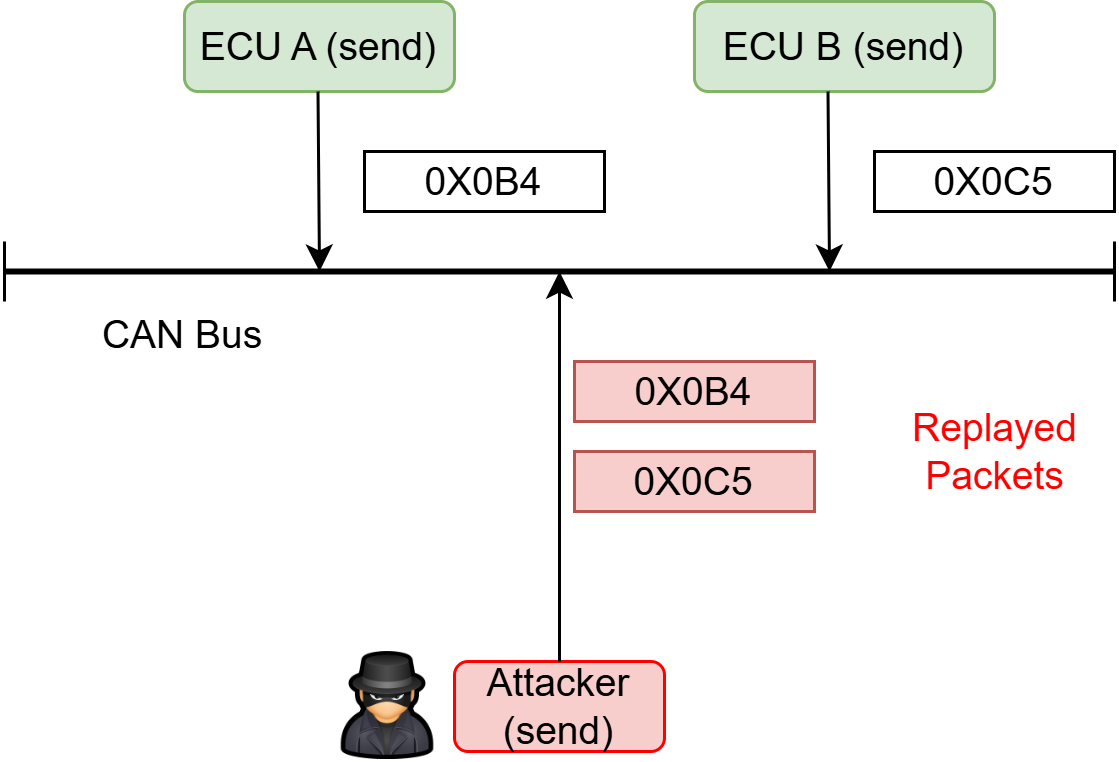}
    \caption{Replay attack}
    \label{fig:replay}
\end{subfigure}
\hfill
\caption{CAN bus attacks}
\label{fig:CAN Bus Attacks}
\end{figure}

\subsubsection{Denial-of-Service (DoS) Attack}

\begin{itemize}
    \item \textbf{Attack Definition:}
The goal of a DoS attack is to overwhelm the CAN bus bandwidth by transmitting large volumes of messages, leading to system malfunctions and service disruptions \cite{rajapaksha2023ai}. Koscher et al.~\cite{koscher2010experimental} demonstrated that DoS attacks can disable individual CAN bus components.

  \item \textbf{Attack Method:} 
Since message priority is determined by the arbitration field, an attacker can exploit the CAN frame priority arbitration scheme vulnerability by sending numerous messages with low CAN IDs (high priority), such as 0x0000. This flooding of high-priority frames occupies the bus, preventing other ECUs from accessing it ~\cite{hoang2022detecting}. Figure ~\ref{fig:flooding} illustrates how a high-priority CAN ID 0x0000 delays a lower-priority CAN ID 0x0B4.

   \item \textbf{Attack Scenario:}
We assume that the attacker has gained access to the in-vehicle network, either physically or remotely, through one of the entry points outlined in Section \ref{entry_points}. Leveraging this access, the attacker floods the CAN bus with high-priority messages, such as those with CAN IDs like 0x0000, without requiring prior knowledge of the CAN bus traffic. The arbitration mechanism prioritises these malicious messages, taking control of the bus and blocking critical communications, such as those from the engine control unit or braking system. For instance, while the vehicle is in motion, an attacker carrying out this attack could disable cruise control or activate emergency braking, preventing critical messages from reaching the appropriate ECU in time and creating potentially hazardous driving conditions. Within seconds, the network's capacity becomes overwhelmed, causing delays that severely compromise vehicle safety. 

\item \textbf{Attack Impact:}
A successful DoS attack not only delays normal messages by occupying the bus \cite{lee2017otids}, but also prevents other ECUs from transmitting frames to the in-vehicle network, significantly impacting network availability \cite{cho2016error}. Such attacks can lead to a complete breakdown of ECU communication and severe disruption of the entire CAN bus network system \cite{hossain2020effective, liu2017vehicle}, posing significant threats to the safety of drivers, passengers, and other road users \cite{fowler2018fuzz}.

\end{itemize}

\subsubsection{Spoofing Attack}

\begin{itemize}
    \item \textbf{Attack Definition:}
In a spoofing attack, an unauthorised attacker targets specific existing CAN IDs and injects fabricated messages to control particular functions. Since CAN IDs appear legitimate, distinguishing between real and spoofed messages becomes challenging, leading to system malfunctions ~\cite{hoang2022detecting}. 

 \item \textbf{Attack Method:}
The attacker may sniff the CAN bus traffic or possess prior knowledge about the ECUs' CAN messages. They can then use this information to inject spoofed messages into the CAN bus. Figure ~\ref{fig:spoofing} illustrates a spoofing attack where an attacker, using the spoofed CAN ID 0x0B4, targets the legitimate CAN ID 0x0B4. This enables the attacker to disrupt vehicle functions by generating manipulated messages that appear legitimate.

 \item \textbf{Attack Scenario:}
We assume that the attacker has gained access to the in-vehicle network, either physically or remotely, through one of the entry points outlined in Section \ref{entry_points}. In this attack, we assume that the attacker has some knowledge of CAN bus traffic by implementing an impersonation attack. One method to achieve this is by connecting a malicious device to eavesdrop on all broadcast traffic, capturing data transmitted across the network. During this reconnaissance phase, the attacker analyses the traffic to identify patterns in ECU behaviour, such as specific CAN IDs, payload structures, and message transmission intervals. Armed with this knowledge, the attacker selects a target ECU, such as the speedometer, with plans to disable it from the bus later. Next, the attacker gains remote access to the internal network and crafts spoofed messages by replicating the target ECU’s CAN ID and injecting false speed readings into the bus. This activity exploits the CAN bus protocol’s error-handling mechanisms, as described in \cite{iehira2018spoofing}. By transmitting dominant bits whenever the legitimate ECU sends recessive bits, the attacker creates intentional bit conflicts. These conflicts generate repeated error frames, eventually causing the legitimate ECU to exceed its error tolerance threshold and enter a "bus-off" state. In the bus-off state, the legitimate ECU is effectively disconnected from the network, unable to send or receive critical messages. The attacker then injects maliciously crafted messages using the same CAN ID as the target ECU. For instance, they could dangerously slow the vehicle on a highway or increase its speed in restricted areas, thereby creating hazardous conditions.

\item \textbf{Attack Impact:}
Spoofing attacks can cause system malfunctions and disrupt vehicle operations \cite{hoang2022detecting}. They pose significant threats to personal safety, particularly when targeting critical ECUs responsible for essential functions such as braking or steering \cite{iehira2018spoofing}.

\end{itemize}

\subsubsection{Frame Fuzzification Attack}

\begin{itemize}
    \item \textbf{Attack Definition:}
The goal of a frame fuzzification attack is to inject random messages into the CAN bus network, making them appear as legitimate traffic. Attackers may exploit prior knowledge of CAN IDs and payload values obtained through CAN bus sniffing, or they may perform the attack without any prior knowledge of CAN frames, treating it as a black-box attack \cite{hossain2020lstm}. In this type of attack, the attacker might alter the CAN ID, the CAN payload, or both simultaneously \cite{aliwa2021cyberattacks}. Since the range of valid CAN packets is relatively small, even simple fuzzing of packets can cause significant damage \cite{koscher2010experimental}.

 \item \textbf{Attack Method:}
The attacker sends arbitrary messages into the CAN bus network, making them appear as legitimate traffic. Figure ~\ref{fig:fuzzuing} illustrates a frame fuzzification attack, where the attacker generates and injects random CAN IDs (e.g. 0x123, 0x357, and 0x222), which are illegitimate.
As a result, all ECUs receive a high volume of functional messages, leading to unintended vehicle behaviours \cite{lee2017otids}. 
For example, Chockalingam et al. \cite{chockalingam2016detecting} introduced Gaussian noise to create a frame fuzzification attack on CAN data. A frame fuzzification attack can disrupt the entire CAN bus network, leading to severe malfunctions such as the steering wheel shaking uncontrollably, signal lights flickering erratically, or automatic and unintended changes in the gear shift \cite{hossain2020effective}.

 \item \textbf{Attack Scenario:}
We assume that the attacker has gained access to the in-vehicle network, either physically or remotely, through one of the entry points outlined in Section \ref{entry_points}. Without prior knowledge of CAN frames, the attacker is able to inject random malicious CAN frames. Using techniques such as fuzzing, the attacker transmits random or malformed messages into the CAN bus to provoke unintended system behaviours or identify exploitable vulnerabilities that could be leveraged in future attacks. Additionally, through reverse engineering, the attacker monitors legitimate traffic to deduce the structure and purpose of CAN messages, enabling the creation of malicious packets to execute specific commands targeting particular ECUs. 


\item \textbf{Attack Impact:}
Frame fuzzification attacks can compromise ECUs, triggering unexpected vehicle behaviours such as steering wheel shaking, erratic signal lights, and unintended gear shifts \cite{lee2017otids}. These behaviours can confuse the driver, potentially resulting in poor decisions or accidents. Such attacks not only disrupt normal vehicle functions but also threaten operational integrity, compromise data privacy, and endanger personal safety, posing significant risks to passengers and other road users.

\end{itemize}

\subsubsection{Replay attack}

\begin{itemize}
    \item \textbf{Attack Definition:}
In replay attacks, attackers store a valid message at a certain time and replay it at a different time without any changes \cite{jo2021survey}. For example, an attacker can store a speedometer reading and later rebroadcast it to the network \cite{al2019intrusion}. In this type of attack, the attacker might alter the ID, payload sequences, or both \cite{rajapaksha2023ai}.

 \item \textbf{Attack Method:}
Attackers first observe and capture valid CAN messages while monitoring the vehicle’s CAN bus. Later, they replay these captured messages without modification to manipulate the system \cite{jo2021survey}. Consequently, the normal CAN packet in the traffic is replaced with a previously captured packet, causing the historical effects \cite{lokman2019intrusion}. Figure ~\ref{fig:replay} illustrates a replay attack, where the attacker sniffs the legitimate CAN messages (e.g., 0X0B4 and 0X0C5), and after some time, he reinjects the same messages into the network.

 \item \textbf{Attack Scenario:}
We assume that the attacker has gained access to the in-vehicle network, either physically or remotely, through one of the entry points outlined in Section \ref{entry_points}. In this attack, we assume that the attacker has no prior knowledge of CAN bus traffic but is able to capture legitimate CAN traffic. One method to achieve this is by connecting a malicious device to sniff the CAN bus traffic, capturing data transmitted across the network. After some time, the attacker gains access to the network and replays the sniffed messages back into the traffic.

\item \textbf{Attack Impact:}
Even though this attack is easy to carry out, as it requires no prior knowledge of traffic operation, it can pose serious safety threats to both vehicles and passengers \cite{rajapaksha2023ai}. 
Koscher et al. \cite{koscher2010experimental} demonstrated replay attacks to manipulate the radio and various body control module functions in the CAN bus. Although the replayed packet is a valid subsequence, the replaced packet disrupts the original packet sequence. Consequently, this can lead to severe issues such as continuous CAN packet transmission requests, deadline violations, and inversion of the CAN arbitration priority scheme. Furthermore, the altered packet sequence prevents the vehicle from functioning properly, as the packets are no longer transmitted sequentially, violating protocol requirements \cite{lokman2019intrusion}.

\end{itemize}

\section{Related Work}
\label{Related_Work}
In this section, we review existing surveys and reviews on IDS approaches in in-vehicle networks, highlighting their contributions and how our survey differs.
There are several surveys in this field. Most of them review the security of in-vehicle networks in general and include IDS as one of the approaches, alongside ML and DL \cite{karopoulos2022demystifying,aliwa2021cyberattacks, dupont2019survey, tomlinson2018towards, al2019intrusion, wu2019survey, rajbahadur2018survey, jo2021survey, lokman2019intrusion, loukas2019taxonomy, young2019survey, quadar2024intrusion, lampe2023intrusion, limbasiya2022systematic}. Others focus specifically on reviewing ML and DL techniques \cite{rajapaksha2023ai, nagarajan2023machine, lampe2023survey, almehdhar2024deep, taslimasa2023security}.

Rajapaksha et al.\cite{rajapaksha2023ai} reviewed and classified state-of-the-art AI-based in-vehicle IDSs, collecting papers using the PRISMA (Preferred Reporting Items for Systematic Reviews and Meta-Analyses) protocol. They proposed an AI-based IDS taxonomy, reviewed benchmark datasets, and outlined the steps for developing AI-based attack detection in the CAN bus. Additionally, they identified and discussed the limitations of current approaches for securing in-vehicle networks and suggested possible future research directions. \\
Karopoulos et al.\cite{karopoulos2022demystifying} compiled a meta-taxonomy that consolidates the key classification features of in-vehicle IDSs proposed in existing surveys, offering a unified perspective on their development. They reviewed available datasets for training and testing in-vehicle IDSs, along with simulators used for dataset generation or performance evaluation. Additionally, they highlighted the main challenges and future directions in this rapidly advancing field. \\
Aliwa et al.\cite{aliwa2021cyberattacks} reviewed cryptographic and IDS solutions to protect vehicular data and discussed their challenges. \\
Dupont et al.\cite{dupont2019survey} categorised in-vehicle IDSs based on the required message count for attack detection, the data utilised, and the design of the detection model. \\
Tomlinson et al.\cite{tomlinson2018towards} reviewed methods for CAN IDSs and categorised them into signature detection and anomaly detection. The anomaly detection category was further divided into statistical, knowledge-based, and ML approaches, highlighting their implications. \\
Al-Jarrah et al.\cite{al2019intrusion} provide an overview of intra-vehicle IDSs, categorizing them into flow-based, payload-based, and hybrid types. They discuss and identify the challenges and current gaps in the landscape of intra-vehicle IDS research. Out of the 42 reviewed papers, only 23 are ML-based IDSs. \\
Wu et al.\cite{wu2019survey} categorised in-vehicle IDSs into four types: fingerprint-based, parameter monitoring-based, information theory-based, and ML-based. Moreover, they discussed the drawbacks and emerging research directions. However, out of the 20 reviewed papers, only 9 focused on ML-based IDSs. \\
Rajbahadur et al.\cite{rajbahadur2018survey} conducted a survey on anomaly detection for securing CAVs. They introduced a taxonomy with three main categories and nine subcategories. In addition, they classified the surveyed papers into 38 dimensions. While the study provided valuable insights, it lacks individual paper summaries and practical implementation strategies. \\
Jo et al.\cite{jo2021survey} classified in-vehicle countermeasures into four categories: preventative protection, IDSs, authentication, and post-protection. They further divided IDS techniques into CAN packet-based and ECU hardware characteristic-based approaches. Although this survey comprehensively examines CAN attack surfaces and corresponding protection mechanisms, it does not focus on ML-based IDSs. \\
Lokman et al.\cite{lokman2019intrusion} introduced a taxonomy for classifying research papers according to four aspects: deployment strategies, attacking techniques, technical challenges, and detection approaches. They also categorised anomaly-based IDSs into frequency-based, ML-based, statistical-based, and hybrid-based. Despite its contribution, only a limited number (five) of works were discussed under the ML-based approach.\\
Loukas et al.\cite{loukas2019taxonomy} provided a comprehensive taxonomy focusing on IDS characteristics and architectures designed for different types of vehicles, such as aircraft, land vehicles, and watercraft. They classified audit techniques into statistical, ML, and rule-based IDSs, and only 13 ML-based IDSs for the CAN bus are discussed. \\
Young et al.\cite{young2019survey} categorised CAN bus IDSs into signature-based and anomaly-based methods. However, the reviewed ML-based IDSs are limited. \\
Lampe and Meng \cite{lampe2023survey} provided a comprehensive overview of DL-based IDSs in automotive networks, categorizing them based on their topologies and techniques, such as DNN-based IDSs, CNN-based IDSs, LSTM-based IDSs, attention- and transformer-based IDSs, and GAN-based IDSs. They also discuss the advantages and disadvantages of each approach. \\
Nagarajan et al.\cite{nagarajan2023machine} presented a comprehensive review of ML-based IDSs for in-vehicle and inter-vehicle communications. They reviewed available datasets, summarised current testbeds, and discussed open research issues. \\
Taslimasa et al.\cite{taslimasa2023security} provided a comprehensive literature review of proposed IDSs for Internet of Vehicles (IoV) networks (inter-vehicle and intra-vehicle) that utilise ML and DL algorithms. Additionally, they discussed IDS criteria, highlighting key factors to consider when assessing IoV network security. \\
Quadar et al.\cite{quadar2024intrusion} reviewed and classified in-vehicle IDSs based on detection methods, including fingerprint-based methods, time- and frequency-based methods, and ML-based methods. However, the reviewed ML-based IDSs are limited. \\
Lampe and Meng \cite{lampe2023intrusion} reviewed automotive intrusion detection methodologies, categorising IDSs into non-learning, traditional ML, and DL. They further classified IDSs by six dimensions: analysis,  deployment,  detection, evaluation, learning, and monitoring modes. Open challenges and future research opportunities were also discussed.
\begin{table}[t!]
    \centering
    \resizebox{\textwidth}{!}{ 
\begin{tabular}{ccccccc}
\hline
\textbf{Reference} & \textbf{Year} & \textbf{Search Strategy} & \textbf{ML-DL Specific} & \textbf{FL-based IDSs} & \textbf{Evaluation metrics} \\ \hline
\cite{tomlinson2018towards} & 2018 &  & $\circ$ &  &  &  \\ \hline
\cite{rajbahadur2018survey} & 2018 & $\bullet$ & $\circ$  &  &   \\ \hline
\cite{al2019intrusion} & 2019 &  & $\circ$ &  & $\bullet$   \\ \hline
\cite{wu2019survey} & 2019 &  & $\circ$ &  &  \\ \hline
\cite{loukas2019taxonomy} & 2019 &  & $\circ$  &  &    \\ \hline
\cite{lokman2019intrusion} & 2019 &  & $\circ$ &  &    \\ \hline
\cite{young2019survey} & 2019 &  & $\circ$ &  &    \\ \hline
\cite{dupont2019survey} & 2019 &  & $\circ$ &  &    \\ \hline
\cite{jo2021survey} & 2021 &  & $\circ$  &  &    \\ \hline
\cite{aliwa2021cyberattacks} & 2021 &  & $\circ$  &  &    \\ \hline
\cite{karopoulos2022demystifying} & 2022 &  & $\circ$ &  &    \\ \hline
\cite{limbasiya2022systematic} & 2022 & $\bullet$ & $\circ$  & &    \\ \hline
\cite{taslimasa2023security} & 2023 &  & $\bullet$  &  & $\bullet$   \\ \hline
\cite{rajapaksha2023ai} & 2023 & $\bullet$ & $\bullet$ &  & $\circ$  \\ \hline
\cite{lampe2023survey} & 2023 &  & $\bullet$  &  & $\circ$   \\ \hline
\cite{nagarajan2023machine} & 2023 &  & $\bullet$ &  & $\circ$     \\ \hline
\cite{lampe2023intrusion} & 2023 &  & $\circ$ &  & $\circ$  \\ \hline
\cite{quadar2024intrusion} & 2024 &  & $\circ$ &  &    \\ \hline
\cite{almehdhar2024deep} & 2024 &  & $\bullet$ & $\circ$ & $\circ$   \\ \hline
\textbf{Our Survey} & 2025 & $\bullet$ & $\bullet$ & $\bullet$ & $\bullet$  \\ \hline
\end{tabular}
}
\begin{tablenotes}
       \centering
       \item \textbf{$\bullet$}: Extensive, \textbf{$\circ$}: Partial \end{tablenotes}
\caption{Comparison with In-Vehicle IDS Surveys}
\label{table:comparison_with_Surveys}
\end{table}

Limbasiya et al. \cite{limbasiya2022systematic} present a systematic survey that extensively analyses different Attack Detection and Prevention System (ADPS) categories for CAVs. They discuss state-of-the-art research in each category, highlighting the latest findings in this domain. Additionally, they identify crucial open security challenges that must be addressed for the secure deployment of CAVs.\\
Almehdhar et al. \cite{almehdhar2024deep} categorised IDS techniques into conventional ML, DL, and hybrid models. They also explored emerging technologies such as FL and Transfer Learning. Key limitations in current methodologies and potential directions for future research were identified.

Even though existing in-vehicle surveys and reviews have made significant contributions to the field, they have certain limitations. As shown in Table \ref{table:comparison_with_Surveys}, few surveys follow a structured search methodology to ensure full coverage and a comprehensive review. In addition, none of these surveys review FL-based IDS, except for the work in \cite{almehdhar2024deep}, which briefly presents some studies in this area. Although Chellapandi et al. \cite{chellapandi2023survey} provide a survey on FL for CAVs, they do not include any in-vehicle IDSs and instead focus on FL applications such as steering wheel angle prediction, vehicle trajectory prediction, object detection, motion control, and driver monitoring. Lastly, although some surveys have reviewed the performance metrics used in the reviewed papers, we provide a comprehensive review of all evaluation metrics, including performance, time, and memory requirements, emphasising the need to include these metrics to develop deployable solutions. \textbf{To the best of our knowledge, this survey is the first to provide a comprehensive review of ML, DL, and FL-based IDS for in-vehicle networks.}
Table \ref{table:comparison_with_Surveys} compares this survey with other existing surveys on in-vehicle IDSs, highlighting the key contributions of this work.

\section{Methodology}
\label{Methodology}
In this section, we outline the search strategy used to collect the reviewed papers.

\begin{figure}[hp!]
\centering
\includegraphics[scale=.20]{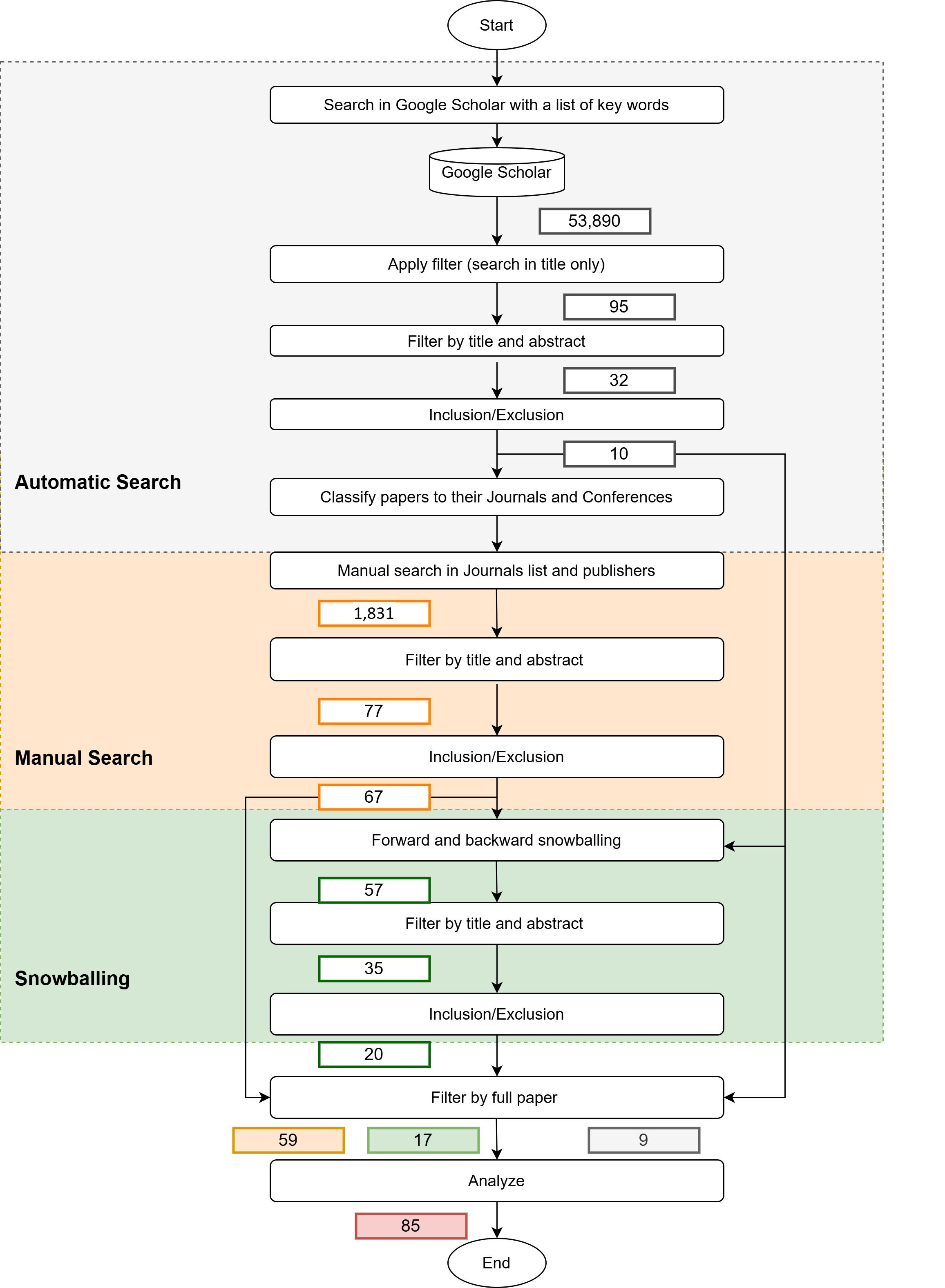}
\caption{Search and selection processes flowchart}
\label{Research Strategy}
\end{figure}

\subsection{Search Strategy}
\label{Search_Strategy}
To select studies for inclusion, we followed Kitchenham’s \cite{kitchenham2013systematic} method, a well-established and effective guide for identifying relevant literature. Although originally designed for the software engineering domain, this method has been widely applied in other fields, including cybersecurity \cite{alhirabi2021security}. Our process began with an automatic search using Google Scholar to minimise bias towards any specific publisher \cite{wohlin2014guidelines} and to identify key publishers and conferences in the field. Based on this search, we compile a list of publishers and conferences for a subsequent manual search. To ensure comprehensive coverage, we employ a snowball approach to locate all related papers. After gathering a substantial number of publications, we applied filtering processes to select the most relevant ones. Finally, we analysed the collected articles (from any time up to and including January 2025) to develop the literature review presented in this paper. Figure \ref{Research Strategy} shows the search strategy adopted.

\subsubsection{Data Sources and Search Strategy}
To begin, we formulated several search queries on Google Scholar using keywords that combine terms representing our area of research and those frequently found in paper keywords. These terms are listed in Table \ref{table:search_terms_results}. Logical operators "AND" and "OR" were utilised to ensure comprehensive results. These queries generated a range of papers, some of which were only loosely relevant. This step also provided insights into the digital libraries and journals that prioritise CAN bus security. Subsequently, we conducted a hybrid search using more complex queries in specific journals and libraries, including IEEE Xplore, Scopus, ACM, MDPI, Springer, and ScienceDirect. Three search strategies were employed during this process: automatic, manual, and snowballing.

\begin{enumerate}
\item \textbf{Automatic search} 
We carried out this stage using the advanced search function in Google Scholar, employing key terms such as "CAN bus", "controller area network", "in-vehicle", "intrusion detection system", "IDS", "anomaly detection", "unknown attacks", and "federated". These root words were chosen because Google Scholar automatically searches for variations of the same word; for instance, searching for "federated" returns results like "federated", "federated learning", "federated-based", and "federated environment". Additionally, the search was not restricted to a specific time period. Initially, using the filter "anywhere in the article," we retrieved an unwieldy number of results (53,890) (see Table \ref{table:search_terms_results}). To refine this, we applied the filter "in the title of the article," reducing the results to 95 papers. Only peer-reviewed papers were included in our analysis.

\item \textbf{Manual search}
In this stage, we applied more complex queries, including specific journals and libraries, such as IEEE Xplore, Scopus, ACM, MDPI, Springer, and ScienceDirect. Table \ref{Queries_Table} lists examples of queries used in the manual search. 

\item \textbf{Snowballing} 
The snowballing technique was applied to the papers identified through automatic and manual searches. This approach included both forward and backward snowballing. Forward snowballing (or citation analysis) locates papers that are cited in the papers found in the initial stages. Backward snowballing (or reference analysis) looks at the reference lists of the papers found in the initial search process. References included in the selected papers were chosen based on a review of the title, abstract, and the paper’s structure. We found that backward snowballing using critical papers was an effective means of identifying relevant papers. The set of articles selected was updated to include any additional relevant studies found by snowballing.
\end{enumerate}

\begin{table}[h!]
\centering
\renewcommand{\arraystretch}{1.2} 
\resizebox{\textwidth}{!}{
\begin{tabular}{p{6cm}cc}
\hline
\textbf{Key Terms} & \textbf{Anywhere in the Article} & \textbf{In the Title} \\
\hline
("CAN bus" OR "Controller Area Network") AND ("IDS" OR "intrusion detection system") & 9,220  & 10  \\
"In-vehicle" AND ("IDS" OR "Intrusion Detection System")  & 20,000  & 25 \\
("CAN bus" OR "Controller Area Network") AND "anomaly detection" & 4,690  & 11 \\
"In-vehicle" AND "anomaly detection" & 9,970 & 39 \\
("CAN bus" OR "Controller Area Network") AND "unknown attacks" &  658 & 0  \\
"In-vehicle" AND "unknown attacks" & 892 & 0  \\
("CAN bus" OR "Controller Area Network") AND "federated" & 1,860 & 0  \\
"In-vehicle" AND "federated" &  6,600 & 10  \\
\hline 
\textbf{Total} & 53,890  &  95 \\ 
\hline                 
\end{tabular}
}
\caption{Google Scholar search terms and results}
\label{table:search_terms_results}
\end{table}

\begin{table}[h!]
\centering
\renewcommand{\arraystretch}{1.5} 
\begin{tabular}{lp{12cm}}
\hline
\textbf{Category} & \textbf{Queries and Terms} \\ \hline
\textbf{General} & 
\begin{tabular}[c]{@{}l@{}}
"CAN bus" OR "controller area network" OR "in-vehicle" AND \\"intrusion detection system" OR "IDS" OR "anomaly detection" OR \\ "unknown attacks" OR "federated".

\end{tabular} \\ \hline

\textbf{More Specific} &
\begin{tabular}[c]{@{}l@{}}

\textbf{ACM:} {[}[Title: "can bus"{]} OR {[}Title: "controller area network"{]} OR \\
{[}Title: "in-vehicle"{]}{]} AND {[}[Title: "intrusion detection system"{]} OR \\
{[}Title: "ids"{]} OR {[}Title: "anomaly detection"{]} OR \\ {[}Title: "unknown attacks"{]} OR {[}Title: "federated"{]}{]} \\ \\

\textbf{IEEE Xplore:} ("Document Title":"CAN bus")\\ OR ("Document Title":"controller area network" ) OR \\ ("Document Title":"in-vehicle") AND \\ ("Document Title":"intrusion detection system") OR \\ ("Document Title":"IDS") OR \\ ("Document Title":"anomaly detection" ) OR \\ ("Document Title":"unknown attacks") OR \\ ("Document Title":"federated") \\ \\

\textbf{Scopus:} ( TITLE-ABS-KEY ( "CAN bus" OR "controller area network"\\ OR "in-vehicle" ) AND TITLE-ABS-KEY ( "intrusion detection system" \\ OR "IDS" OR "anomaly detection" OR "unknown attacks" OR \\"federated" ) ) \\ \\


\end{tabular} \\ \hline
\end{tabular}
\caption{Examples of queries and terms used for the online library search}
\label{Queries_Table}
\end{table}

\subsubsection{Selection Strategy}
\begin{figure}[H]
\centering
\includegraphics[scale=.70]{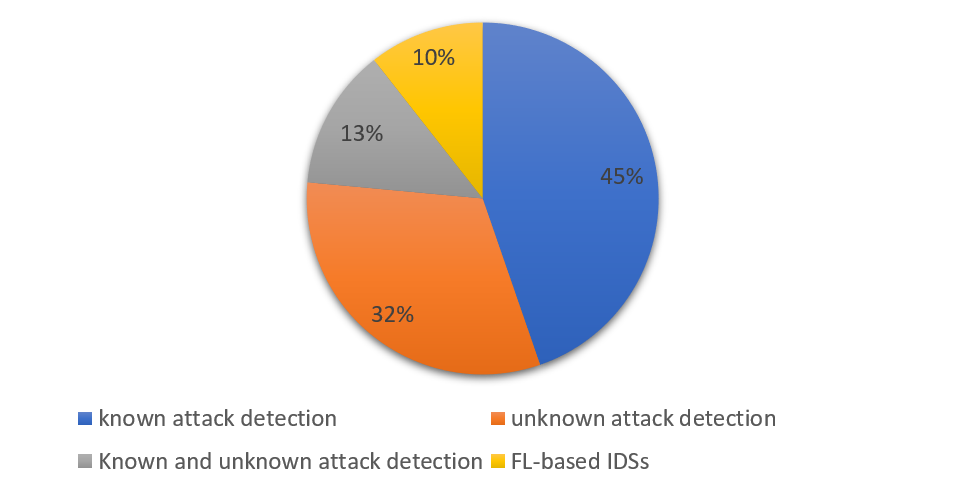}
\caption{Distribution of collected papers by category}
\label{collected_papers_chart}
\end{figure}

The selection strategy involved defining inclusion and exclusion criteria, as well as applying filtering during the search process. The steps in the selection process included applying the inclusion and exclusion criteria, followed by an additional filtering stage involving a quality assessment to ensure the selection of high-quality studies. Each of these steps is discussed below.
\begin{itemize}
    \item \textbf{Filtering Irrelevant Papers}: The papers collected through manual, automatic, and snowballing approaches included several that did not apply directly to our study and had to be eliminated. Elimination was conducted in two steps. In the first, papers were excluded based on the title, keywords, abstract, and sometimes the conclusion in case of any doubts. Based on this step, a decision was made regarding whether to include each paper in the next step. Eliminating was undertaken after each search (automatic, manual, and snowballing) to reduce the number of papers. When a paper passed the initial elimination step, it was subjected to the inclusion and exclusion criteria.

    \item \textbf{Inclusion and Exclusion Criteria}
    In this process, we defined our exclusion and inclusion criteria. A paper was considered for exclusion if it met one or more criteria. The exclusion criteria for each paper were as follows: (i) not written in English; (ii) was a review or survey paper; (iii) lacked a full version (e.g., only a poster or abstract); (iv) did not employ an ML or DL approach; (v) required reverse engineering; (vi) used other data alongside CAN bus data; and (vii) employed other approaches such as statistical, rule-based, or physical fingerprinting methods. Papers that were not excluded were then evaluated according to an inclusion list of other criteria. If no inclusion criteria were met, the paper was rejected. The inclusion criteria were as follows: (i) focused on the CAN bus protocol rather than other in-vehicle protocols; and (ii) focused on attack detection as the primary goal. Non-peer-reviewed papers were included only if they were strictly relevant to the topic and had a high citation rate, or if the author was well-known in the field.

\end{itemize}
As shown in Figure \ref{Research Strategy}, reading the full paper is the final step in the search and selection process, filtering out the collected papers from the previous steps. The papers from the automatic search were reduced from 10 to 9 after filtering by reading the full text. In the manual search within journal lists and publishers, the initial 1,831 papers were filtered down to 59. For snowballing, 57 papers were initially selected and reduced through filters to 17. Thus, the total number of collected papers comprises 9 from the automatic search, 59 from the manual search, and 17 from snowballing, resulting in a total of 85 papers for analysis. Of these, 38 focused on known attack detection, 27 on unknown attack detection, 11 on IDSs capable of detecting both known and unknown attacks, and 9 on FL-based IDSs. Figure \ref{collected_papers_chart} illustrates the number of collected papers in each category, highlighting that known attack detection is the most researched area, while significantly less work has been conducted on IDSs capable of detecting both known and unknown attacks, as well as on FL-based IDSs.

\section{Intrusion Detection Systems for In-Vehicle Networks}
\label{Existing_IDSs}
\begin{figure}[H]
\centering
\includegraphics[scale=.13]{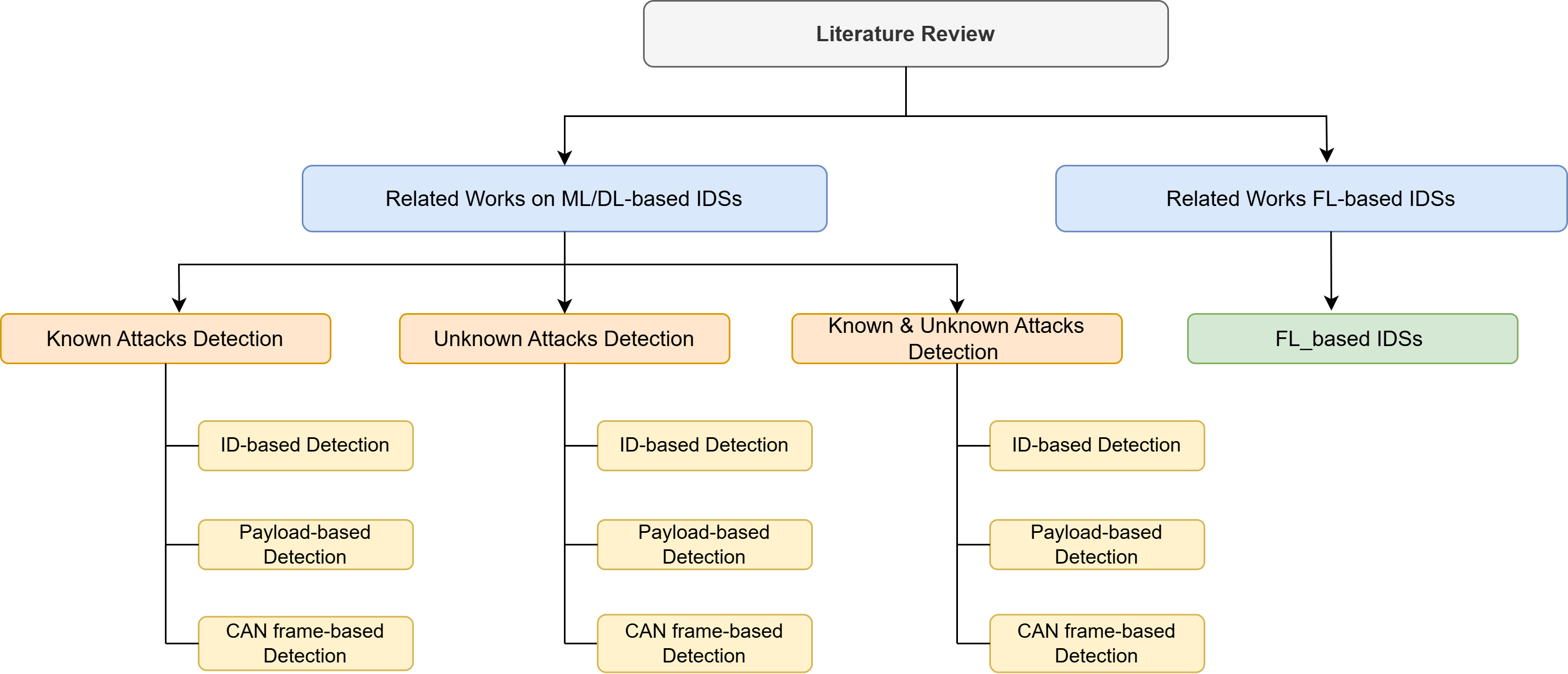}
\caption{Categories of reviewed literature}
\label{categories_chart}
\end{figure}

In this section, we begin by introducing IDSs for in-vehicle networks and highlighting the differences between in-vehicle IDSs and those used for other applications. We then provide an overview of in-vehicle IDS approaches.  Additionally, we categorise the collected papers into three categories: known attack detection, unknown attack detection, and work capable of detecting both known and unknown attacks for analysis. Figure \ref{categories_chart} illustrates the categories and subcategories of the reviewed literature. Lastly, we review all the evaluation metrics used in the reviewed papers.

\subsection{Intrusion Detection System for In-Vehicle Networks}
According to NIST SP 800-94, intrusion detection is “the process of monitoring the events occurring in a computer system or network and analysing them for signs of possible incidents” \cite{hernandez2023intrusion}. Therefore, an IDS is typically considered a software or hardware system designed to automatically detect suspicious activity in a network ~\cite{young2019survey}. In vehicle networks, IDSs are crucial for identifying malicious attacks ~\cite{aliwa2021cyberattacks}. They can be implemented as either host-based or network-based systems ~\cite{young2019survey}. Host-based IDSs are installed on each vehicle's ECU, allowing comprehensive monitoring of internal ECU operations. In contrast, network-based IDSs are deployed within the CAN network or central gateways to oversee all network traffic. However, Host-based IDSs are not a viable solution for vehicles, as they require a change in ECUs that are not cost effective ~\cite{aliwa2021cyberattacks}. In contrast, deploying a network-based IDS as an additional node on the CAN bus is a more feasible and practical solution, as it avoids the need for any CAN bus modifications ~\cite{rajapaksha2023ai}. Unlike IDSs in other applications, in-vehicle IDSs are constrained by computing power, memory size, and communication capabilities. This is because modern ECUs in vehicles are primarily powered by 32-bit embedded processors, with limited computational performance and memory resources ~\cite{wu2019survey}.

\subsection{Overview of In-Vehicle IDS Approaches}
Research on developing in-vehicle IDSs has grown significantly in recent years, driven by the critical need to enhance the security of in-vehicle networks and detect cyberattacks. Researchers have explored various approaches to building these systems. IDSs can be classified as either signature-based, for detecting known attacks, or anomaly-based, for identifying new, unknown attacks ~\cite{hoppe2009applying}. Anomaly-based IDSs are further categorised into statistical, ML, rule-based, and physical fingerprinting methods ~\cite{rajapaksha2023ai}. 
However, this work specifically focuses on the development of in-vehicle IDSs using ML and DL approaches. This focus arises from the widespread use of ML and DL-based IDSs to process large volumes of CAN traffic data. These approaches efficiently extract and pre-process raw CAN data, which is critical as vehicle manufacturers often do not provide detailed specifications for decoding these raw data ~\cite{rajapaksha2023ai}. This section is divided into three categories: known attack detection, unknown attack detection, and work capable of detecting both known and unknown attacks.

\subsection{Known Attacks Detection}
\label{Known_Attacks_Detection}
\begin{figure}[p]  
\centering
    \rotatebox{90}{\includegraphics[width=0.75\paperheight, height=0.8\paperwidth]{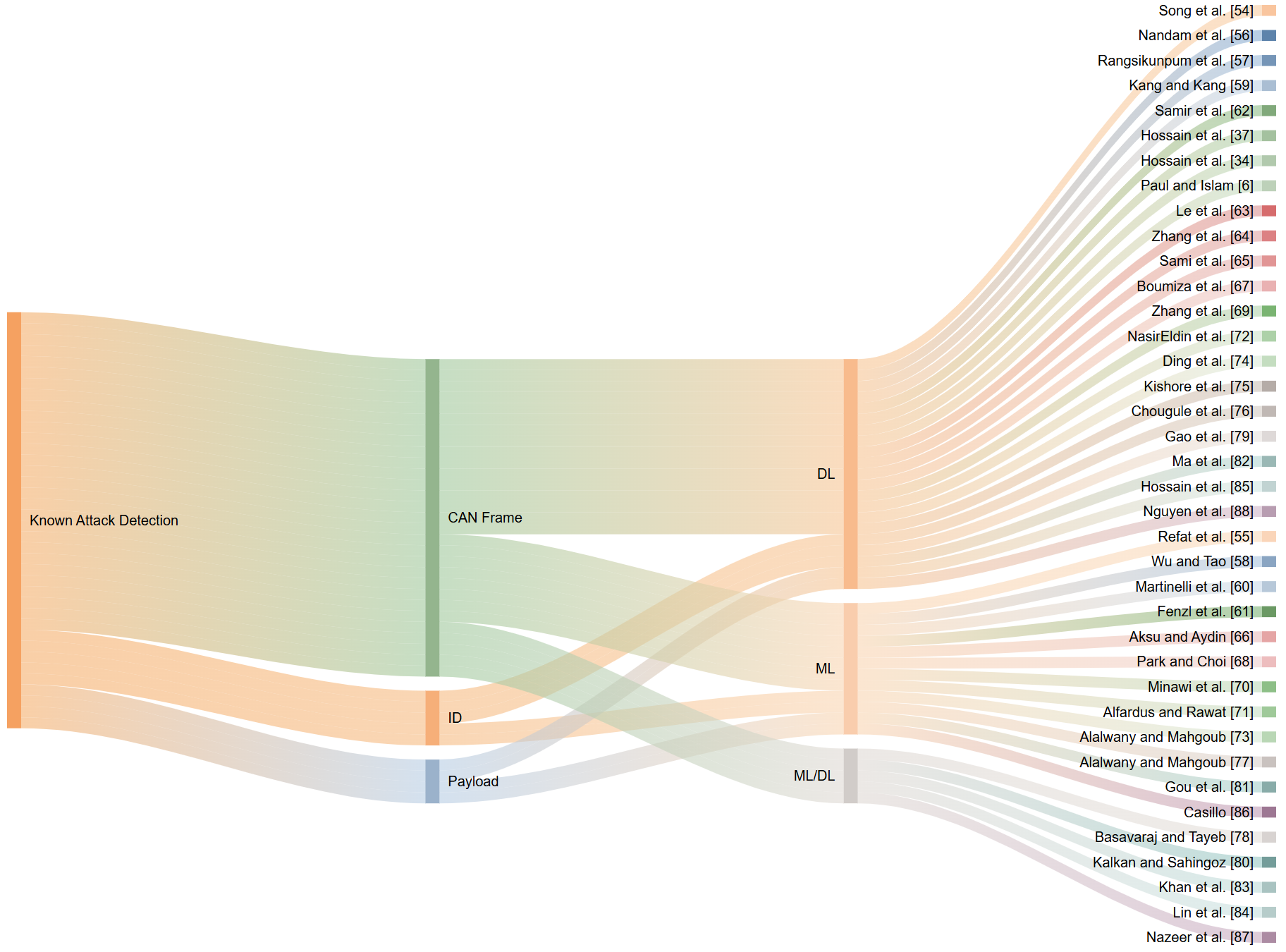}}  
    \caption{Related work on known attack detection}
    \label{fig:known_Sankey_Diagram}
\end{figure}

As mentioned in Section \ref{Search_Strategy}, there are 38 papers on IDSs focusing on known attack detection. In this section, we analyse these papers and discuss the existing methodologies used to detect or classify known threats in in-vehicle networks. Detection of known attacks typically relies on supervised learning, where models are trained on labelled data. The section is organised into three subsections based on the features used to build the model: ID-based detection, payload-based detection, and CAN frame-based detection. Each subsection examines different approaches for identifying malicious activities, emphasizing their strengths and limitations. Figure \ref{fig:known_Sankey_Diagram} illustrates previous work on detecting known attacks, showing that most studies utilised a DL approach and used CAN frames as input features.


\subsubsection{ID-Based Detection} 
Attacks such as injecting or deleting frames alter certain properties of message ID sequences compared to normal messages. This section presents research where the authors utilised these properties and used CAN IDs solely as an input feature to develop IDSs for detecting known attacks.

Song et al.~\cite{song2020vehicle} utilised the sequential behaviour of CAN data to identify message injection attacks. During these attacks, frequent frame injections resulted in distinct ID pattern changes, which were leveraged for detection. The authors relied solely on the bit-wise CAN ID sequence, which was processed directly as input, eliminating the need for additional feature engineering. They introduced a deep convolutional neural network (DCNN) model that was redesigned by minimizing unnecessary complexities within the Inception-ResNet architecture to achieve an optimised input size of (29 × 29 × 1) and a binary classification output.

Refat et al.~\cite{refat2022detecting} used graph-based techniques to extract features from the CAN IDs in in-vehicle networks. The authors converted a window of CAN IDs into a graph and extracted seven graph properties, including the number of nodes, number of edges, radius, diameter, density, reciprocity, average clustering coefficient, and assortative coefficient, to use as input features. These extracted features were then used to train two traditional ML algorithms: support vector machine (SVM) and k-nearest neighbours (KNN) models. The experimental results demonstrated that using graph-based features outperformed the traditional CAN bus features.

Nandam et al. \cite{nandam2022can} employed an Long Short-Term Memory (LSTM) model to detect DoS attacks on the CAN bus. The model utilises the CAN ID of incoming messages to identify potential DoS attacks. A sequence of previous messages is stored and combined with the current message to form the input, enabling the model to predict and detect DoS attacks effectively.

Rangsikunpum et al. \cite{rangsikunpum2024fpga} introduced a Binarised Neural Network (BNN)-based IDS to identify attacks on the CAN bus. The primary objective of the proposed IDS is to deploy a ML model on a low-cost Field Programmable Gate Array (FPGA) device, optimising for low power consumption, minimal execution time, and high accuracy. By leveraging a 1-bit BNN model, the implementation is resource-efficient, making it suitable for deployment on cost-effective FPGA devices with reduced power requirements. Moreover, the IDS employs a two-stage architecture: the first stage identifies the presence of an attack, and the second stage, triggered only upon detecting an attack, performs detailed attack classification.

Wu and Tao \cite{wu2024network} proposed a model based on ensemble learning using a Stacking integration approach. The method incorporates a meta-classifier composed of DTs, Extra Trees (ET), and extreme gradient boosting (XGBoost). Final classification predictions are made by linearly combining input features and weights through a SoftMax meta-learner. Ensemble learning in this approach utilises the prediction results as new features, along with the true labels, to train the meta-learner.

\subsubsection{Payload-Based Detection}

Some attacks, such as spoofing, use legitimate CAN IDs but modify the CAN payload values, depending on the characteristics of the particular attack. These changes alter the pattern of payload sequences. This section discusses IDSs that utilise this property and use the CAN payload as an input feature to detect known attacks.

Kang and Kang ~\cite{kang2016intrusion} developed a deep neural network (DNN)-based IDS to defend against malicious attacks. The authors utilised the 8-byte CAN payload to extract features, employing mode and value information to achieve dimensionality reduction. The initial weights for the DNN model were obtained from a separate Deep Belief Network. They then applied a template-matching method to compare the training samples with new CAN packets for detecting malicious messages. Although the proposed model demonstrated improved detection performance, it relied on mode and value information from CAN data, which presents significant challenges, especially without access to the DBC file.

Similarly, Martinelli et al. ~\cite{martinelli2017car} utilised the eight CAN payload features to assess whether these features can effectively discriminate between attacks and normal messages. To address this question, they employed four fuzzy classification algorithms to identify four types of attacks targeting the CAN bus, including DoS, fuzzy, RPM, and gear spoofing. These algorithms include two types of fuzzy-rough KNN, the discernibility classifier, and a fuzzy unordered rule induction algorithm. The classification analysis was performed using the Weka3 tool. Experimental results indicated that the feature vector is a potential candidate for accurately classifying between injected and normal messages.

Fenzl et al. ~\cite{fenzl2021vehicle} used decision trees (DTs) trained through genetic programming (GP) to detect intrusions in the CAN bus. Their approach focuses solely on message payloads, with models trained individually for each CAN ID within the CAN bus training data. The authors compared their method with artificial neural networks (ANNs). Experimental results showed that for most intrusions, the accuracy of the ANN was slightly higher, and the ANN had a significantly lower training time; however, the proposed GP method demonstrated significantly improved detection time.

Samir et al. \cite{samir2024machine} investigated two DL-based IDSs: one leveraging LSTM and the other utilising a one-dimensional convolutional neural network (CNN). These two supervised learning algorithms serve as classifiers capable of categorising attacks into different types. The authors employed two public datasets and a new dataset that they manually generated using the ICSim simulation tool to cover more complex scenarios and attack types. Experimental results show that the LSTM-based IDS outperforms the CNN-based IDS, leveraging its ability to capture temporal patterns for robust detection of diverse CAN bus attacks.

\subsubsection{CAN Frame-Based Detection}
Rather than relying solely on CAN IDs or CAN payload as features, IDSs in the literature have utilised a combination of features to identify pattern changes in CAN data sequences. This approach offers the advantage of detecting alterations in CAN IDs and manipulations of the payload. This section reviews IDSs that use CAN IDs and payload as input features to detect known attacks. Some studies also incorporate the DLC feature or time differences between consecutive CAN IDs, in combination with CAN ID and payload.

Hossain et al. ~\cite{hossain2020lstm} proposed an LSTM-based IDS. The model considers both CAN ID, DLC, and payload as input features to detect point and contextual anomalies. The proposed LSTM is trained on both benign and attack data and employs both binary and multi-class classification. The authors collected CAN messages from an actual Toyota hybrid car and generated three attack scenarios, including DoS, fuzzy, and spoofing attacks. They compared the performance of the proposed LSTM method with the survival analysis method and found that the LSTM model achieves a higher detection rate than the other methods.

Following their earlier research, Hossain et al. ~\cite{hossain2020effective} introduced a 1D CNN model as an alternative to the LSTM model proposed in their previous study. They collected normal datasets from three cars: Toyota, Subaru, and Suzuki, and injected anomalous frames to create attacks, including DoS, fuzzy, RPM, and gear spoofing. The proposed model achieved a high attack detection rate for all types of attacks. However, they considered fuzzy to be the most critical attack in the in-vehicle system, as it is difficult to detect due to its similarity to legitimate traffic within the CAN bus network.

Similarly, Paul and Islam ~\cite{paul2021artificial} proposed an ANN-based anomaly detection method to identify unauthorised messages in CAN bus. They utilised benign and attack classes from DoS and fuzzy datasets to train their ANN model. The model demonstrated a high detection accuracy in distinguishing between legitimate and anomalous messages, achieving nearly negligible rates of false positives and false negatives.

Le et al. \cite{le2024multi} proposed an IDS for multiclass classification based on a combination of AE models and a time-embedded transformer. The AE-based packet-level extraction model learns a compressed representation of each CAN frame within a CAN sequence, while the time-embedded transformer, which replaces positional encoding with a timestamp encoding component, is used as the sequence extraction component.

Zhang et al. \cite{zhang2024efficient} introduced a Binarized CNN (BCNN)-based IDS, designed to leverage the temporal and spatial characteristics of CAN messages. The proposed IDS consists of two main components: an input generator and a BCNN model. The input generator converts CAN messages from feature vectors into image form, enabling the BCNN model to capture their temporal and spatial features. The second component employs the BCNN model to process the output images from the input generator. Experimental results demonstrated that the BCNN model is four times faster and requires less memory compared to a 32-bit CNN-based IDS.

Sami et al. \cite{sami2020rapid} introduced the Network Embedded System Laboratory's IDS (NESLIDS), which employs a supervised DL algorithm based on a DNN. NESLIDS is designed as an anomaly detection system to identify three known attacks.

Aksu and Aydin \cite{aksu2022mga} proposed a meta-heuristic algorithm, the Modified Genetic Algorithm (MGA), to select a subset of features by removing irrelevant ones, thereby improving classification performance and reducing dimensionality. They evaluated the effectiveness of the feature selection process using five classifiers: Support Vector Classifier (SVC), Logistic Regression Classifier (LRC), Decision Tree Classifier (DTC), k-Nearest Neighbors Classifier (KNC), and Linear Discriminant Analysis Classifier (LDAC).

Boumiza et al. \cite{boumiza2019anomaly} proposed an IDS for the CAN bus based on a Multi-Layer Perceptron (MLP) neural network. The IDS first partitions data by the ID field of CAN packets, using the K-means clustering algorithm to create subclusters. It then extracts mode and frequency features from each subcluster to train the neural network. The proposed IDS operates separately for each CAN ID, combining the individual decisions to calculate a final score and trigger an alert in the event of an attack detection.

Park and Choi \cite{park2020hierarchical} proposed a multi-labeled hierarchical classification (MLHC) IDS to detect message injection attacks. MLHC identifies the occurrence of attacks and classifies them using only pre-existing labeled attack data. The authors evaluated the method's performance using four ML algorithms: SGD, kNN, DT, and RF. Simulation results showed that the MLHC model achieved high accuracy with the RF algorithm and rapid detection with the DT algorithm.

Zhang et al. \cite{zhang2020convolutional} proposed a Convolutional Encoder Network (CEN) model designed to detect network intrusions in CAN networks. The architecture integrates an encoder for dimensionality reduction, a CNN to increase network depth, and Inception ResNet to optimise training time. Additionally, the authors introduced a Feature-based Sliding Window method to extract features from the CAN Data Field and CAN IDs. Experimental results highlight the effectiveness of the feature-based sliding window in improving detection performance.

Minawi et al. \cite{minawi2020machine} proposed an ML-based IDS system comprising three layers: the CAN Message Input Layer, the Threat Detection Layer, and the Alert Layer. The Threat Detection Layer utilises ML algorithms such as Random Tree (RT), Random Forest (RF), Stochastic Gradient Descent (SGD) with hinge loss, and Naive Bayes (NB) to detect different types of attacks. Additionally, this layer is designed with multiple modules, each tailored to detect specific types of attacks.

similarly, Alfardus and Rawat \cite{alfardus2021intrusion} used the same proposed IDS in \cite{minawi2020machine} but with four different ML algorithms, including KNN, RF, SVM, and Multilayer Perceptron (MLP), to detect CAN bus attacks.

NasirEldin et al. \cite{nasreldin2021vehicle} proposed an attention-based model to detect CAN bus intrusions. The model consists of an attention layer that assigns higher importance to the most prominent features by calculating attention scores between the input features and the target, followed by a self-attention layer to identify relationships between data elements. Experimental results demonstrate that the proposed model outperformed baseline models, including an LSTM.

Alalwany and Mahgoub \cite{alalwany2022classification} proposed an ML-based IDS for detecting attacks on the CAN bus using supervised ML models, including RF, DT, Gaussian Naïve Bayes (GaussianNB), Logistic Regression (LR), AdaBoost, KNN, XGBoost, and Gradient Boosting. To further enhance attack detection accuracy, the authors combined all supervised models using three ensemble methods: voting, stacking, and bagging. The ensemble learning strategy offers the advantage of enabling models with different capabilities to complement one another in the classification task. Compared to individual models, the ensemble classifiers outperformed the supervised classifiers, improving the effectiveness of the supervised ML models by leveraging diverse learning mechanisms to support one another.

Ding et al. \cite{ding2022vehicle} proposed an IDS based on a Bidirectional LSTM (Bi-LSTM) network with a sliding window strategy. A two-dimensional input data sample set was constructed using the sliding window, and the Bi-LSTM network was trained on these features to learn a classifier for intrusion detection. Experimental results demonstrate that the proposed model outperforms other network models, except for DoS attacks.

similarly, Kishore et al. \cite{kishore2024intelligent} proposed a Bi-LSTM, which processes input data in both forward and backward orientations to detect anomalies in the CAN bus.

Chougule et al. \cite{chougule2024hybridsecnet} proposed HybridSecNet, a hybrid two-step LSTM-CNN IDS designed to enhance in-vehicle security. HybridSecNet consists of two classification stages: the first stage uses an LSTM to classify input data as either normal or attacked. If an attack is detected in the initial stage, the second stage is activated, employing a CNN-based multiclass classifier to identify and categorise the specific type of attack.

Moreover, Alalwany and Mahgoub \cite{alalwany2024effective} proposed an in-vehicle IDS to improve the accurate detection and classification of CAN bus attacks in real time using ensemble techniques and the Kappa Architecture. The Kappa Architecture facilitates real-time attack detection, while ensemble learning combines multiple ML classifiers, including RF, DT, and XGBoost, to enhance detection accuracy. The study demonstrated that ensemble approaches, which integrate the strengths of multiple models, significantly improved detection accuracy and robustness.

Basavaraj and Tayeb \cite{basavaraj2022towards} proposed a lightweight DNN-based model to detect and classify attacks on the CAN bus. The proposed model outperformed baseline models, including RF, DTs, and the kNN algorithm.

Gao et al. \cite{gao2023multi} proposed a CNN and Bi-LSTM model with multi-head attention for attack detection and classification. The CNN module enhances feature extraction, the Bi-LSTM module captures sequential features and relationships, and the multi-head attention module identifies further correlations between features.

Kalkan and Sahingoz \cite{kalkan2020vehicle} applied six different ML models: RF, bagging, ADA boosting, NB, LR, and ANN. Their experimental results demonstrated that tree-based and ensemble learning algorithms achieved superior performance. However, the authors did not specify the features used for training, leading to the assumption that all features were included.

Gou et al. \cite{gou2023multi} proposed an adaptive tree-based ensemble network (ATBEN) as the intrusion detection engine for IDS in the IoV. ATBEN leverages a variety of ML models, including XGBoost, LightGBM, RF, and ET,  as base estimators, stacking them into layers within the network. The cascading connections between layers facilitate precise and efficient multiclass classification. The authors demonstrated the effectiveness of the proposed IDS by evaluating its performance against a range of cyberattacks targeting both in-vehicle systems and external networks within the IoV.

Ma et al. \cite{ma2022gru} introduced a lightweight IDS for the CAN bus, leveraging a GRU-based architecture. To enhance efficiency, they employed a low-complexity feature extraction algorithm to derive features from CAN frames. The proposed model demonstrated near real-time performance and outperformed baseline models in detection accuracy.

Khan et al. \cite{khan2024divacan} proposed DivaCAN, an IDS that combines DL models with conventional ML methods through an ensemble of base classifiers, including DNN, MLP, light gradient-boosting machines, ET, RF, Bagging, and KNN to detect intrusions on the CAN bus. To improve detection performance, a meta-classifier aggregates the outputs of the base classifiers in a weighted and adaptive manner, considering their performances and correlations. This work addresses the trade-off between false positives and time complexity in CAN bus IDS.

Lin et al. \cite{lin2022using} proposed a CNN-based approach leveraging the VGG16 classifier to learn attack behaviour characteristics and classify threats. Feature vectors were transformed into feature images, which were then input into the VGG16 model for accurate categorisation of cyber threats in in-vehicle networks. To ensure high precision in predicting the stability of network intrusion detection, the approach combines the VGG16 model with the XGBoost ensemble learning algorithm, enabling effective analysis of suspicious network traffic.

Hossain et al. \cite{hossain2020long} proposed an LSTM-based IDS for detecting in-vehicle attacks. The CAN message data was collected using a tool called Vehicle Spy 3. To evaluate the IDS, the authors employed both binary and multi-class classification approaches, utilizing vanilla LSTM and stacked LSTM models. Since the dataset originally contained no attacks, the authors simulated DoS, Fuzzing, and Spoofing attacks on the CAN bus of a Toyota Hybrid car using a Python-based program.

Casillo \cite{casillo2019embedded} proposed an embedded IDS for automotive systems by adopting Bayesian Networks for the rapid identification of malicious messages on the CAN bus. The CAN bus dataset was generated by simulating vehicle driving for approximately 24 hours on a city track within the CARLA environment. During the simulation, the vehicle was subjected to attacks to replicate potential intrusion scenarios based on specific use cases.

Nazeer et al. \cite{nazeer2024enhancing} proposed a hybrid approach, DeepXG, which combines the XGBoost and DNN models to detect and classify attacks on the CAN bus. The XGBoost model is trained on the dataset to extract critical features and reduce computational complexity, while the DNN leverages these learned representations to detect anomalies and intrusions.

Nguyen et al. \cite{nguyen2023transformer} proposed an IDS based on a Transformer attention network for a CAN bus, designed to analyse a single message. The proposed IDS includes two models: one using only a single message and another leveraging sequential CAN IDs. The first model effectively detects DoS, fuzzy, and spoofing attacks but cannot detect replay attacks due to its reliance on single-message analysis. To address this limitation, the second model was designed to detect replay attacks by incorporating sequential CAN ID information. Additionally, the proposed model employs transfer learning to enhance the performance of models trained on small datasets from other car models.

Table \ref{Known_attack_detection} summarises the related work on known attack detection methods, including the learning approach, binary or multi-class classification, dataset used, detectable attacks, employed algorithm, and the model size or the size of trainable parameters. In Table \ref{Known_attack_detection}, we assume that papers that do not explicitly state the input features used are referring to CAN frame features. Among these studies, only three \cite{song2020vehicle, fenzl2021vehicle, le2024multi} measure the trainable parameters, reflecting the model's size, while the others did not consider model size for deployment.

\begin{table}[H]
\resizebox{\textwidth}{!}{
\begin{tabular}{ccccccccccc}
\hline
\textbf{Reference} & \textbf{Year} & \textbf{ML / DL} & \textbf{Category} & \textbf{\begin{tabular}[c]{@{}c@{}}Classification \\ Type \end{tabular}} & \textbf{Dataset} & \textbf{ID} & \textbf{Payload} & \textbf{Attack Types} & \textbf{Algorithm} & \textbf{\begin{tabular}[c]{@{}c@{}}Model Size/\\  Trainable Parameters\end{tabular}} \\ \hline
\multicolumn{11}{c}{\cellcolor[HTML]{dcdcdc}\textbf{ID-Based Attack Detection}} \\ \hline
\cite{song2020vehicle} & 2020 & DL & Supervised & Binary & Car Hacking \cite{seo2018gids} & \checkmark &  & Message Injection & DCNN & 1.76 Million \\ \hline
\cite{refat2022detecting} & 2022 & ML & Supervised & Binary & Car Hacking \cite{seo2018gids} & \checkmark &  & \begin{tabular}[c]{@{}c@{}}DoS, Fuzzy, \\ RPM Spoofing\end{tabular} & SVM, KNN & N/A\\ \hline
\cite{nandam2022can} & 2022 & DL & Supervised & Binary & car-hacking   \cite{song2020vehicle} & \checkmark &  & DoS & LSTM & N/A \\ \hline
\cite{rangsikunpum2024fpga} & 2024 & DL & Supervised & \begin{tabular}[c]{@{}c@{}}Binary / \\  Multi-class\end{tabular} & Car Hacking \cite{seo2018gids} & \checkmark &  & \begin{tabular}[c]{@{}c@{}}(Gear, RPM) Spoofing, \\  DoS, Fuzzy\end{tabular} & BNN &  4.85 Mb \\ \hline
\cite{wu2024network} & 2024 & ML & Supervised & Binary & Car Hacking \cite{seo2018gids} & \checkmark &  & \begin{tabular}[c]{@{}c@{}}(Gear, RPM) Spoofing, \\  DoS, Fuzzy\end{tabular} & \begin{tabular}[c]{@{}c@{}}Ensemble model \\ (DT, ET , XGBoost)\end{tabular} & N/A \\ \hline
\multicolumn{11}{c}{\cellcolor[HTML]{dcdcdc}\textbf{Payload-Based Attack Detection}} \\ \hline
\cite{kang2016intrusion} & 2016 & DL & Supervised & Binary & Simulation & \textbf{} & \checkmark & Injection Attacks & DNN & N/A\\ \hline
\cite{martinelli2017car} & 2017 & ML & Supervised & Binary & Car Hacking \cite{seo2018gids} & \textbf{} & \checkmark & \begin{tabular}[c]{@{}c@{}}DoS, Fuzzy, \\(Gear, RPM) Spoofing \end{tabular} & KNN & N/A\\ \hline
\cite{fenzl2021vehicle} & 2021 & ML & Supervised & Binary & \begin{tabular}[c]{@{}c@{}}Car Hacking \cite{seo2018gids}, Tesla\\  Model X data, Renault \\ Zoe electric car data\end{tabular} & \textbf{} & \checkmark & (RPM, Gear) Spoofing  & DT, GP  & 1,101 \\ \hline
\cite{samir2024machine} & 2024 & DL & Supervised & Multi-class & \begin{tabular}[c]{@{}c@{}}Car Hacking\cite{seo2018gids},\\  OTIDS \cite{lee2017otids}, Own\end{tabular} & \textbf{} & \checkmark & \begin{tabular}[c]{@{}c@{}}DoS, Fuzzy, \\ Spoofing,  Replay\end{tabular} & CNN, LSTM & N/A \\ \hline
\multicolumn{11}{c}{\cellcolor[HTML]{dcdcdc}\textbf{CAN Frame -Based Attack Detection}} \\ \hline
\cite{hossain2020lstm} & 2020 & DL & Supervised & \begin{tabular}[c]{@{}c@{}}Binary /  \\ Multi-class\end{tabular} & Own & \checkmark & \checkmark & \begin{tabular}[c]{@{}c@{}}DoS, Fuzzy,\\ Spoofing\end{tabular} & LSTM & N/A\\ \hline
\cite{hossain2020effective} & 2020 & DL & Supervised & \begin{tabular}[c]{@{}c@{}}Binary / \\ Multi-class\end{tabular} & Own & \checkmark & \checkmark & \begin{tabular}[c]{@{}c@{}}DoS, Fuzzy, \\ Spoofing\end{tabular} & CNN & N/A\\ \hline
\cite{paul2021artificial} & 2021 & DL & Supervised & Binary & OTIDS \cite{lee2017otids} & \checkmark & \checkmark & DoS, Fuzzy & ANN & N/A\\ \hline
\cite{le2024multi} & 2024 & DL & Supervised & Multi-class & \begin{tabular}[c]{@{}c@{}}car-hacking \cite{song2020vehicle},\\  ROAD \cite{verma2020road}\end{tabular} & \checkmark & \checkmark & \begin{tabular}[c]{@{}c@{}}(Gear, RPM) Spoofing, \\  DoS, Fuzzy, Fabrication,\\  Masquerade\end{tabular} & \begin{tabular}[c]{@{}c@{}}AE, Time-embedded \\ Transformer\end{tabular} & 259,000 \\ \hline
\cite{zhang2024efficient} & 2024 & DL & Supervised & Binary & Own & \checkmark & \checkmark & Replay, Spoofing & BCNN & N/A \\ \hline
\cite{sami2020rapid} & 2020 & DL & Supervised & Binary & \begin{tabular}[c]{@{}c@{}}OTIDS \cite{lee2017otids}, \\  ML350 \cite{ML3502019Dataset}\end{tabular} & \checkmark & \checkmark & \begin{tabular}[c]{@{}c@{}}DoS, Fuzzy,\\   Impersonation\end{tabular} & DNN & N/A \\ \hline
\cellcolor[HTML]{FFFFFF}\cite{aksu2022mga} & \cellcolor[HTML]{FFFFFF}2022 & \cellcolor[HTML]{FFFFFF}ML & \cellcolor[HTML]{FFFFFF}Supervised & \cellcolor[HTML]{FFFFFF}\begin{tabular}[c]{@{}c@{}}Binary / \\  Multi-class\end{tabular} & \cellcolor[HTML]{FFFFFF}\begin{tabular}[c]{@{}c@{}}car-hacking  \cite{song2020vehicle}\end{tabular} & \cellcolor[HTML]{FFFFFF}\checkmark & \cellcolor[HTML]{FFFFFF}\checkmark & \begin{tabular}[c]{@{}c@{}}(Gear, RPM) Spoofing, \\  DoS, Fuzzy\end{tabular} & \cellcolor[HTML]{FFFFFF}\begin{tabular}[c]{@{}c@{}}SVC, LRC, DTC,\\  KNC, LDAC\end{tabular} & N/A \\ \hline
\cite{boumiza2019anomaly} & 2019 & DL & Supervised & Binary & Dataset \cite{taylor2018probing} & \checkmark & \checkmark & \begin{tabular}[c]{@{}c@{}}Frequency  modification,\\ Data-content  modification\end{tabular} & MLP & N/A \\ \hline
\cite{park2020hierarchical} & 2020 & ML & Supervised & \begin{tabular}[c]{@{}c@{}}Binary / \\  Multi-class\end{tabular} & \begin{tabular}[c]{@{}c@{}}Survival Analysis Dataset \cite{han2018anomaly}\end{tabular} & \checkmark & \checkmark & \begin{tabular}[c]{@{}c@{}}Fuzzy, Flooding,\\  Malfunction\end{tabular} & \begin{tabular}[c]{@{}c@{}}SGD, kNN, \\ DT, RF\end{tabular} & N/A \\ \hline
\cite{zhang2020convolutional} & 2020 & DL & Supervised & Multi-class & \begin{tabular}[c]{@{}c@{}}Car Hacking \cite{seo2018gids},\\   car-hacking \cite{song2020vehicle}\end{tabular} & \checkmark & \checkmark & \begin{tabular}[c]{@{}c@{}}(Gear, RPM) Spoofing, \\  DoS, Fuzzy\end{tabular} & CEN & N/A \\ \hline
\cite{minawi2020machine} & 2020 & ML & Supervised & Binary & Car Hacking \cite{seo2018gids} & \checkmark & \checkmark & \begin{tabular}[c]{@{}c@{}}(Gear, RPM) Spoofing, \\  DoS, Fuzzy\end{tabular} & RT, RF, SGD, NB & N/A \\ \hline
\cite{alfardus2021intrusion} & 2021 & ML & Supervised & Binary & Car Hacking \cite{seo2018gids} & \checkmark & \checkmark & \begin{tabular}[c]{@{}c@{}}(Gear, RPM) Spoofing, \\  DoS, Fuzzy\end{tabular} & \begin{tabular}[c]{@{}c@{}}KNN, RF, \\  SVM, MLP\end{tabular} & N/A \\ \hline
\cite{nasreldin2021vehicle} & 2021 & DL & Supervised & Binary & Car Hacking \cite{seo2018gids} & \checkmark & \checkmark & \begin{tabular}[c]{@{}c@{}}(Gear, RPM) Spoofing, \\  DoS, Fuzzy\end{tabular} & Attention-based model & N/A \\ \hline
\cite{alalwany2022classification} & 2022 & ML & {\color[HTML]{222222} Supervised} & Binary & \begin{tabular}[c]{@{}c@{}}Car Hacking: Attack \& Defence \\  Challenge 2020  \cite{kang2021car}\end{tabular} & \checkmark & \checkmark & \begin{tabular}[c]{@{}c@{}}Flooding, Spoofing,\\   Replay, Fuzzy\end{tabular} & \begin{tabular}[c]{@{}c@{}}LR, GaussianNB, \\ k-NN, RF,  Gradient Boosting, \\ AdaBoost, DT, XGBoost\end{tabular} & N/A \\ \hline
\cite{ding2022vehicle} & 2022 & DL & Supervised & Binary & Car Hacking \cite{seo2018gids} & \checkmark & \checkmark & \begin{tabular}[c]{@{}c@{}}(Gear, RPM) Spoofing, \\  DoS, Fuzzy\end{tabular} & Bi-LSTM & N/A \\ \hline
\cite{kishore2024intelligent} & 2024 & DL & Supervised & Binary & \begin{tabular}[c]{@{}c@{}}Car Hacking: Attack \& Defence \\  Challenge 2020  \cite{kang2021car}\end{tabular} & \checkmark & \checkmark & \begin{tabular}[c]{@{}c@{}}Flooding, Spoofing,\\   Replay, Fuzzy\end{tabular} & Bi-LSTM & N/A \\ \hline
\cite{chougule2024hybridsecnet} & 2024 & DL & Supervised & \begin{tabular}[c]{@{}c@{}}Binary / \\  Multi-class\end{tabular} & Car Hacking \cite{seo2018gids} & \checkmark & \checkmark & \begin{tabular}[c]{@{}c@{}}DoS, Fuzzy,\\   (Gear, RPM) Spoofing\end{tabular} & LSTM-CNN & N/A \\ \hline
\cite{alalwany2024effective} & 2024 & ML & {\color[HTML]{222222} Supervised} & Multi-class & \begin{tabular}[c]{@{}c@{}}Car Hacking: Attack \& Defence \\  Challenge 2020  \cite{kang2021car}\end{tabular} & \checkmark & \checkmark & \begin{tabular}[c]{@{}c@{}}DoS, Spoofing, \\  Replay, Fuzzy\end{tabular} & RF, DT,  XGBoost & N/A \\ \hline
\cite{basavaraj2022towards} & 2022 & DL / ML & {\color[HTML]{222222} Supervised} & Multi-class & CAN dataset \cite{Dupont2019Dataset} & \checkmark & \checkmark & \begin{tabular}[c]{@{}c@{}}Reconnaissance, \\  DoS, Fuzzy\end{tabular} & DNN & N/A \\ \hline
\cite{gao2023multi} & 2023 & DL & Supervised & Multi-class & Car Hacking \cite{seo2018gids} & \checkmark & \checkmark & \begin{tabular}[c]{@{}c@{}}DoS, (Gear, RPM) \\ Spoofing, Fuzzy\end{tabular} & CNN, bi\_LSTM & N/A \\ \hline
\cite{kalkan2020vehicle} & 2020 & ML / DL & Supervised & Binary & Car Hacking \cite{seo2018gids} & \checkmark & \checkmark & \begin{tabular}[c]{@{}c@{}}(Gear, RPM) \\ Spoofing, \\  DoS, Fuzzy\end{tabular} & \begin{tabular}[c]{@{}c@{}}RF, bagging, \\ ADA boosting, \\ NB, LR, ANN\end{tabular} & N/A \\ \hline
\cite{gou2023multi} & 2023 & ML & {\color[HTML]{222222} Supervised} & Multi-class & \begin{tabular}[c]{@{}c@{}}car-hacking   \cite{song2020vehicle}\end{tabular} & \checkmark & \checkmark & \begin{tabular}[c]{@{}c@{}}(Gear, RPM) Spoofing, \\  DoS, Fuzzy\end{tabular} & \begin{tabular}[c]{@{}c@{}}XGBoost,  LightGBM,\\   RF,  ET\end{tabular} & N/A \\ \hline
\cite{ma2022gru} & 2022 & DL & Supervised & Binary & Car Hacking \cite{seo2018gids} & \checkmark & \checkmark & \begin{tabular}[c]{@{}c@{}}DoS, Spoofing,\\  Fuzzy\end{tabular} & GRU & N/A \\ \hline
\cite{khan2024divacan} & 2024 & ML / DL & Supervised & Multi-class & OTIDS \cite{lee2017otids} & \checkmark & \checkmark & \begin{tabular}[c]{@{}c@{}}DoS, Fuzzy, \\  Impersonation\end{tabular} & \begin{tabular}[c]{@{}c@{}}DNN, MLP, \\ light gradient-boosting \\ machine, ET, RF,  \\ Bagging, KNN\end{tabular} & N/A \\ \hline
\cite{lin2022using} & 2022 & DL / ML & Supervised & {\color[HTML]{222222} Multi-class} & Car Hacking \cite{seo2018gids} & \checkmark & \checkmark & \begin{tabular}[c]{@{}c@{}}(Gear, RPM) \\ Spoofing, \\  DoS, Fuzzy \end{tabular} & VGG16, XGBoost & N/A \\ \hline
\cite{hossain2020long} & 2020 & DL & Supervised & \begin{tabular}[c]{@{}c@{}}Binary / \\  Multi-class\end{tabular} & Own & \checkmark & \checkmark & \begin{tabular}[c]{@{}c@{}}DoS, Fuzzy,  \\  Spoofing\end{tabular} & LSTM & N/A \\ \hline
\cite{casillo2019embedded} & 2019 & ML & Supervised & Binary & Simulation & \checkmark & \checkmark & \begin{tabular}[c]{@{}c@{}}Turn right, \\  Turn left, Brake\end{tabular} & Bayesian Network & N/A \\ \hline
\cite{nazeer2024enhancing} & 2024 & ML / DL & Supervised & Multi-class & Own & \checkmark & \checkmark & \begin{tabular}[c]{@{}c@{}}Flooding, Replay, \\ Spoofing\end{tabular} & XGBoost, DNN & N/A \\ \hline

\cite{nguyen2023transformer} & 2023 & DL & Supervised & \begin{tabular}[c]{@{}c@{}}Binary / \\  Multi-class\end{tabular} & \begin{tabular}[c]{@{}c@{}}Car Hacking \cite{seo2018gids},\\  IVN \cite{HCRL2019Dataset}, \\  Survival Analysis Dataset \cite{han2018anomaly}\end{tabular} & \checkmark & \checkmark & \begin{tabular}[c]{@{}c@{}}(Gear, RPM) Spoofing, \\  DoS, Fuzzy, \\ Replay, Malfunction \end{tabular} & Transformer & N / A \\ \hline
\end{tabular}
} 
\caption{Summary of related work on known attack detection methods}
\label{Known_attack_detection}
\end{table}

\subsubsection{Limitations of Known Attacks Detection}
\label{Limitations_of_Known_Attacks_Detection}
Despite the high accuracy and low false alarm rate (FAR) of the proposed known attack detection models, their performance heavily depends on well-labelled attack data and balanced datasets. However, obtaining labelled data remains a significant challenge for researchers \cite{hernandez2023intrusion}. Moreover, the labeling process is often time-consuming, prone to errors, and tedious \cite{said2020network}.

Additionally, the main limitation of these studies is that none of the models are capable of detecting new attacks or deviations from the known attacks they were trained on. Because attackers continuously attempt to evade detection and use new, previously unseen attacks, supervised learning-based models struggle to recognise unfamiliar attack patterns not present in the training data \cite{vikram2020anomaly}. This limitation can lead to significant security consequences.

\subsection{Unknown Attacks Detection}
\label{Unknown_Attacks_Detection}
\begin{figure}[p]  
\centering
    \rotatebox{90}{\includegraphics[width=0.75\paperheight, height=0.8\paperwidth]{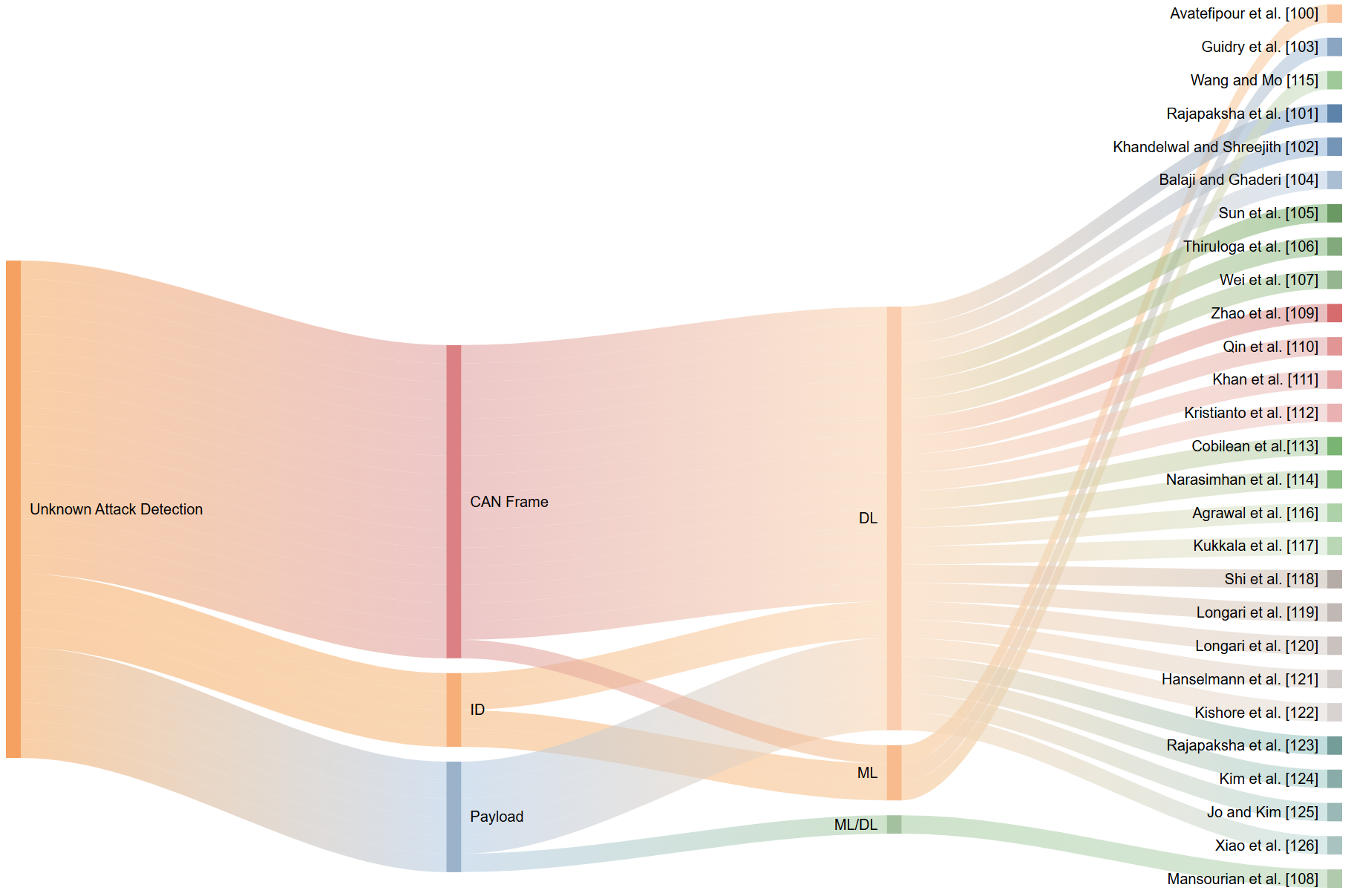}}  
   \caption{Related work on unknown attack detection}
    \label{fig:unknown_Sankey_Diagram}
\end{figure}

As mentioned in Section \ref{Search_Strategy} there are 27 papers on IDSs focusing on unknown attack detection or anomaly detection. In this section, we analyse these papers and discuss the existing methodologies used to detect new, unknown threats in in-vehicle networks. Detection of unknown attacks typically relies on unsupervised learning, where models are trained solely on normal data, relying on profiling normal traffic behaviours to detect anomalous traffic that could indicate a potential attack. As a result, unsupervised learning-based models are well suited to detecting previously unseen attacks \cite{pratomo2018unsupervised}. The section is organised into three subsections based on the features used to build the model: ID-based detection, payload-based detection, and CAN frame-based detection. Each subsection examines different approaches for identifying malicious activities, emphasizing their strengths and limitations. Figure \ref{fig:unknown_Sankey_Diagram} illustrates previous work on detecting unknown attacks. As depicted in the figure, similar to known attack detection, most studies on unknown attack detection utilised a DL approach and used CAN frames as input features.

\subsubsection{ID-Based Detection}
This section reviews research where authors used only CAN IDs as the input feature to develop IDSs for detecting new, unknown attacks.

Avatefipour et al. \cite{avatefipour2019intelligent} proposed an anomaly detection model based on a modified one-class support vector machine (OCSVM) incorporating the modified bat algorithm (MBA). The model was built using normal CAN bus traffic, which exhibits recurring patterns in CAN IDs under normal conditions. Any deviation from this normal traffic, such as increased message occurrence frequency or message flooding, is detected by the model as malicious activity. For evaluation, CAN bus data were collected from a licensed, unmodified vehicle and two public CAN bus datasets. The authors compared the proposed model with baseline Isolation Forest and classical OCSVM models, finding that the MBA-OCSVM achieved the highest true positive rate and the lowest false alarm rate compared to both alternatives.

Rajapaksha et al. \cite{rajapaksha2022keep} proposed CAN-CID, a context-aware IDS aimed at addressing the computational inefficiency of N-gram-based models while detecting a wide range of cyberattacks on the CAN bus. CAN-CID utilises an ensemble approach that combines a Gated Recurrent Unit (GRU) network and a time-based model. The single-layer GRU network detects anomalous ID sequences and minimises detection latency, while the time-based model identifies anomalies using time-based thresholds. The anomaly-to-total-ID ratio within an observation window is then used to classify the window as either anomalous or benign. This study highlights the effectiveness of ensemble models in detecting diverse attacks on the CAN bus.

Khandelwal and Shreejith \cite{khandelwal2023real} presented a convolutional autoencoder (CAE) model for detecting zero-day attacks, trained solely on benign CAN messages. Leveraging Vitis-AI tools, they quantised the model to optimise performance on resource-constrained platforms. The proposed IDS achieves state-of-the-art classification accuracy across multiple unseen attacks, along with a 1.3x speed-up in processing latency and approximately 2x reduction in power consumption compared to existing state-of-the-art IDSs.

Guidry et al. \cite{guidry2023one} proposed the use of a One-Class Support Vector Machine (OC-SVM) to detect anomalous data on a vehicle's CAN bus. Instead of utilising raw CAN bus data, three distinct features were extracted for each unique CAN ID: the average frequency of appearance of a CAN ID, the average time interval between consecutive appearances of a CAN ID, and the standard deviation of transmission times for CAN IDs. These features were selected because they rely on the temporal and behavioural characteristics of message transmissions rather than the data content within the messages. The model was trained on CAN bus data collected under normal operating conditions, making it well-suited for detecting unknown attacks in vehicular networks.

\subsubsection{Payload-Based Detection}
This section discusses IDSs that use the 8-byte CAN payload as an input feature to identify new, unknown attacks.

Balaji and Ghaderi \cite{balaji2021neurocan} proposed NeuroCAN, a contextual anomaly detection model that consists of an embedding layer and LSTM to learn the spatio-temporal correlations among CAN payload values. The embedding layer performs a linear transformation of the input data from each CAN ID, passes it through a sigmoid function, and accumulates it over all IDs. This is followed by an LSTM and an output layer, forming a prediction-based anomaly detector. The use of payload values from other IDs as context enables the capture of inter-ID correlations. However, the model is trained separately for each CAN ID, resulting in high memory and computational costs.

Sun et al. \cite{sun2021anomaly} proposed a CNN-LSTM-based IDS with an attention mechanism. The model used one-dimensional convolution to extract abstract features, while a bi-directional LSTM was employed to capture time dependencies. The bit flip rate was used to identify continuous fields from the 64-bit payload, resulting in a 41-bit smaller signal, which is more efficient than directly predicting the full 64-bit. Experiments demonstrated that this approach reduces data dimensionality and improves model training efficiency. The pre-processed data was fed into the neural network model to predict the output signal and determine whether the received signal was abnormal. The proposed model improved attack detection accuracy by 2.5\% compared to related research.

Thiruloga et al. \cite{thiruloga2022tenet} introduced TENET, a novel anomaly detection framework based on temporal convolutional neural attention (TCNA) networks. TENET takes a sequence of signal values from a message as input and uses CNNs to predict the signal values of the next message instance by learning the underlying probability distribution of normal data. A DT-based classifier was then employed as the attack detector. Experimental results showed that TENET achieved a 3.32\% improvement in detection accuracy, a 32.7\% reduction in the false negative rate, and 94.62\% fewer model parameters compared to a baseline model. However, the model processed data ID-wise, training separate models for each ID, which limits its ability to detect anomalies, such as collective anomalies, that arise from interactions between different CAN IDs.

Wei et al. \cite{wei2022novel} introduced AMAEID, a multi-layer denoising autoencoder model. The model takes only the 8-byte payload of the CAN message as input. It first transforms the raw hexadecimal payload into binary format, then applies a multi-layer denoising autoencoder to extract deeper hidden features that represent the underlying characteristics of the message. Additionally, AMAEID utilises an attention mechanism and a fully connected layer to classify messages as normal or abnormal. Experimental results demonstrate that AMAEID surpasses traditional ML algorithms like DT, KNN, and LinearSVC. However, the model was trained and tested using only two CAN IDs.

Mansourian et al. \cite{mansourian2023anomaly} proposed an anomaly-based IDS to detect attacks on the in-vehicle CAN bus. The proposed IDS comprises three modules: an LSTM model, a prediction error calculator, and a Gaussian Naïve Bayes (GNB) classifier. The LSTM is trained on normal CAN messages to learn the typical sequential behaviour of each ECU. Once trained, the network predicts the next expected payload of an ECU based on past observations and compares it to the actual received value. When an attack occurs, the trained LSTM network fails to make accurate predictions, resulting in higher-than-normal prediction errors. The GNB classifier then classifies messages as either normal or an attack based on these prediction errors.

Zhao et al. \cite{zhao2022gvids} introduced the Same Origin Method Execution (SOME) attack, which mimics the period, clock skew, and voltage of normal messages, making detection by existing IDSs challenging. To address this, they developed a GAN-based IDS, named GVIDS, which employs one-hot encoding to represent data and converts data frames into CAN images. This approach is effective as attacks either directly alter frame data or disrupt frame sequences, indirectly modifying all the consecutive data fields. Experiments on two real vehicles show that GVIDS successfully detects SOME attacks as well as other existing attack types.

\subsubsection{CAN Frame-Based Detection}
This section reviews IDSs that use the CAN frame (CAN IDs and payload) as input features to detect new, unknown attacks. Some studies also incorporate the DLC feature and/or time differences between consecutive CAN IDs, in combination with CAN ID and payload.

Qin et al. \cite{qin2021application} proposed an LSTM-based anomaly detection algorithm to detect abnormal behaviour on the CAN bus. CAN data were collected from the network of a real vehicle, with simulated attacks such as tampering or inserting duplicate random packets into the CAN bus. CAN ID and payload were converted from hexadecimal to binary representations instead of decimal, which increased the dimensionality of the features. Anomaly detection in the message stream of the CAN bus was performed for each ID separately. Experimental results showed that the proposed model could detect anomalous data with over 90\% accuracy.

Khan et al. \cite{khan2021enhanced} used a bidirectional LSTM model with an improved feature processing technique to address the challenge of zero-day attacks. The proposed IDS is a multi-stage system, where the initial stage employs a state-based Bloom filter technique to verify the states of incoming data, while the second stage uses a bidirectional LSTM classifier to detect cyberattacks. They applied enhanced data pre-processing to improve the scalability and performance efficiency of the IDS, including feature conversion, feature reduction, and feature normalization. Principal component analysis was used for feature reduction. Experimental results showed that the feature pre-processing led to a 19.31\% improvement in accuracy compared to raw data.

Kristianto et al. \cite{kristianto2024sustainable} proposed a lightweight unsupervised IDS on a simple Recurrent Neural Network (RNN). The authors suggest deploying the IDS model at each domain gateway, leveraging the computational resources of the gateways to handle only domain-specific messages. This approach enables the gateway to be optimised for detecting malicious messages within its domain while maintaining a lightweight design. The IDS achieves up to a 94\% reduction in parameters compared to existing models, significantly decreasing memory usage and energy consumption. Despite this reduction in size, the proposed models demonstrate only a slight decrease in accuracy compared to current solutions.

Cobilean et al. \cite{cobilean2023anomaly} proposed a Transformer neural network-based IDS designed to predict anomalous behaviour within CAN protocol communication. The Transformer model is trained to predict the next communication sequence, and anomalies are detected when the difference between the predicted sequence and the actual received sequence exceeds a defined threshold. A key advantage of this model is that it does not require labelled attack data for learning the communication sequence.

Narasimhan et al. \cite{narasimhan2021unsupervised} proposed an unsupervised two-stage approach that combines DL with a probabilistic model for anomaly detection. In the first stage, an autoencoder (AE) is used to extract optimal features that differentiate between normal data and attacks on the CAN bus. Unlike other autoencoder-based models that utilise the reconstructed signal for anomaly detection, this model leverages the latent space as input to a Gaussian Mixture Model (GMM). In the second stage, the GMM clusters these features into normal and attack categories. Experimental results demonstrated that the proposed method achieved superior performance across various datasets. For evaluation, a real dataset from a Mercedes ML350 was used; however, as this dataset contained only four CAN IDs, the practical applicability of the model may be constrained.

Wang and Mo \cite{wang2021can} proposed a CAN bus anomaly detection model based on the FLXGBoost algorithm. To address the challenge posed by the large volume of traffic data messages with limited features, they introduced a newly defined feature: information entropy, which serves as an additional set of features in the CAN message data domain.

Agrawal et al. \cite{agrawal2022novelads} proposed NovelADS, an IDS that utilises CNNs and LSTMs to detect anomalies in CAN network traffic. NovelADS captures spatio-temporal features and long-term dependencies from CAN messages. The DL models are trained on normal CAN data, and the system classifies incoming CAN data as genuine or anomalous using a reconstruction-based thresholding approach.

Kukkala et al. \cite{Kukkala2020INDRA} proposed INDRA, an IDS based on a GRU-based recurrent autoencoder designed to learn latent representations of normal CAN traffic and detect anomalies on the CAN bus. At runtime, the trained autoencoder monitors deviations from normal behaviour to identify potential intrusions. Signal-level intrusion scores, calculated as the difference between predicted and actual signal values, are used to identify anomalous signals. The authors trained separate autoencoder models for each CAN ID, enabling ID-specific anomaly detection model.

Shi et al. \cite{shi2024ids} introduced an IDS called IDS-DEC, which integrates a spatiotemporal self-encoder employing LSTM and CNN (LCAE) with an entropy-based deep embedding clustering approach. The LSTM component models the sequential nature of the data, capturing long-term dependencies in the time-series data from the CAN bus. Additionally, as network data can be represented as a multidimensional matrix with spatial structure, CNNs are employed to extract key features, thereby enhancing the accuracy and efficiency of detection. Experimental results demonstrate that the proposed IDS achieves superior detection performance compared to traditional ML algorithms and other deep clustering methods.

Longari et al. \cite{longari2020cannolo} introduced CANnolo, an IDS based on LSTM autoencoders for identifying anomalies on the CAN bus. CANnolo analyses CAN message streams to construct a model of normal data sequences and detects anomalies by measuring the discrepancy between reconstructed sequences and their corresponding real sequences. The authors partitioned the dataset into groups based on CAN IDs, with each group processed independently and trained on separate models. While this approach simplifies the training process, it limits the system's ability to detect signal correlations, thereby reducing its effectiveness in identifying anomalies such as collective anomalies \cite{rajapaksha2023ai}.

To improve the overall architecture and reduce the computational requirements of CANnolo, ensuring it meets the real-time constraints of the automotive domain, Longari et al. \cite{longari2023candito} then proposed CANdito, an unsupervised IDS that leverages LSTM autoencoders to detect anomalies using a signal reconstruction process. CANdito reconstructs the time series of CAN packets for each ID and calculates anomaly scores based on the reconstruction error.

Hanselmann et al. \cite{hanselmann2020canet} proposed CANet, an LSTM-based autoencoder designed to identify attacks on the CAN bus. Separate LSTM models were used for each CAN ID, with their outputs concatenated into a single latent vector. The difference between the original and reconstructed signal values was utilised to determine the normal status. Experimental results showed that the model achieved a high detection rate with low false-positive and false-negative rates across various attack types.

Similarly, Kishore et al. \cite{kishore2022deep} proposed an LSTM-based anomaly detection method. The model outperforms previous tree-based ML algorithms, including AdaBoost, GBoost, Bagging, XGBoost, and LGBM.

Rajapaksha et al. \cite{rajapaksha2023beyond} introduced an ensemble IDS that combines a GRU network and a novel AE model called Latent AE to identify cyberattacks on the CAN bus.The GRU network analyses the CAN ID field, while Latent AE focuses on the CAN payload field to identify anomalies. To improve efficiency, Latent AE incorporates Cramér’s statistic-based feature selection and a transformed CAN payload structure. By utilising a compact latent space, it overcomes the issue of high false negatives in traditional AEs caused by overgeneralisation. Experimental findings reveal that the ensemble IDS enhances attack detection and addresses the limitations of the individual models.

Kim et al. \cite{kim2023anomaly} proposed an IDS based on multiple LSTM-Autoencoders that utilise diverse features, including transmission intervals and payload value changes, to capture various characteristics of normal network behaviour. The system consists of a feature sequence extractor, LSTM-Autoencoder models, and an anomaly detector. The time interval sequence extractor calculates the intervals between consecutive frames with the same ID, generating a chronological sequence for each ID. Similarly, the Hamming distance sequence extractor computes the Hamming distances between the payloads of consecutive frames within ID-based streams. These feature sequences are processed by the LSTM-Autoencoders to produce reconstructed sequences. The anomaly detector evaluates the differences between the original time interval and Hamming distance sequences and their reconstructed counterparts, using these differences to determine whether the frame sequences are normal or anomalous.

Jo and Kim \cite{jo2024intrusion} proposed an IDS based on the Transformer architecture, which predicts the next data point based on the flow of previously input data. The IDS can detect attacks affecting both the temporal and spatial aspects of the data, as CAN data comprise temporal information recorded over time and spatial information recorded across devices. This two-dimensional data is used to train the model, achieving higher performance compared to using one-dimensional data.

Xiao et al. \cite{xiao2019robust} proposed an anomaly detection IDS for in-vehicle networks based on a Convolutional LSTM Network (ConvLSTM), which accounts for both temporal and spatial correlations. The ConvLSTM model is first trained on benign CAN data, and its predictions are used to calculate the correlation coefficient with actual data. Abnormal behaviour is detected by comparing the correlation coefficients between the predicted and real data. Experimental results indicate that the ConvLSTM model maintains a stable correlation coefficient for normal data, while the coefficient for attack data declines rapidly over time. Compared to the LSTM model, the ConvLSTM model more effectively captures the underlying features of benign data, producing a more consistent correlation coefficient for attack-free states. Furthermore, the sharp drop in the correlation coefficient for attack data can facilitate the detection of unknown attacks.

Table \ref{Unknown_attack_detection} summarises the related work on unknown attack detection methods, including the learning approach, dataset used, detectable attacks, employed algorithm, and the model size or the size of trainable parameters. In Table \ref{Unknown_attack_detection}, we assume that papers that do not explicitly state the input features used are referring to CAN frame features. 

Among these studies, only three \cite{song2020vehicle, fenzl2021vehicle, le2024multi} measure the trainable parameters, reflecting the model's size, while the others did not consider model size for deployment.

\begin{table}[H]
\resizebox{\textwidth}{!}{
\begin{tabular}{cccccccccc}
\hline
\textbf{Reference} & \textbf{Year} & \textbf{ML / DL} & \textbf{Category} & \textbf{Dataset} & \textbf{ID} & \textbf{Payload} & \textbf{Attack Types} & \textbf{Algorithm} & \textbf{\begin{tabular}[c]{@{}c@{}}Model Size/\\  Trainable Parameters\end{tabular}} \\ \hline
\multicolumn{10}{c}{\cellcolor[HTML]{dcdcdc}\textbf{ID-Based Attack Detection}} \\ \hline
\cite{avatefipour2019intelligent} & 2019 & ML & Unsupervised & \begin{tabular}[c]{@{}c@{}}Own,\\ Dodge \cite{DodgeCANMessages}, \\ OTIDS \cite{lee2017otids}\end{tabular} &  \checkmark &  & Injection & MBA-OCSVM  & N/A\\ \hline
\cite{rajapaksha2022keep} & 2022 & DL & Unsupervised & \begin{tabular}[c]{@{}c@{}}ROAD \cite{verma2020road}, \\ car-hacking \cite{song2020vehicle}, \\  Survival Analysis Dataset \cite{han2018anomaly}\end{tabular} & \checkmark &  & \begin{tabular}[c]{@{}c@{}}Fabrication, \\  Suspension,\\  Masquerade\end{tabular} & GRU & N/A \\ \hline
\cite{khandelwal2023real} & 2023 & DL & Unsupervised &  car-hacking   \cite{song2020vehicle}  & \checkmark &  & \begin{tabular}[c]{@{}c@{}}DoS, Fuzzy, \\  (Gear, RPM) Spoofing\end{tabular} & AE & N/A \\ \hline
\cite{guidry2023one} & 2023 & ML & Unsupervised & Own & \checkmark &  & \begin{tabular}[c]{@{}c@{}}Random ID, \\  Zero ID, Replay\end{tabular} & OC-SVM & N/A \\ \hline
\multicolumn{10}{c}{\cellcolor[HTML]{dcdcdc}\textbf{Payload-Based Attack Detection}} \\ \hline
\cite{balaji2021neurocan} & 2021 & DL & Unsupervised &  Two Public Datasets from \cite{seo2018gids} & \textbf{} & \checkmark & \begin{tabular}[c]{@{}c@{}}Flood, Replay, \\  Drop, Spoofing,\\   Fuzzy\end{tabular} & LSTM & N/A \\ \hline
\cite{sun2021anomaly} & 2021 & DL & Unsupervised & \begin{tabular}[c]{@{}c@{}}CAN Signal\\  Extraction and \\ Translation \cite{song2020discovering}\end{tabular} & \textbf{} & \checkmark & \begin{tabular}[c]{@{}c@{}}Flood, Replay, \\   Drop, Spoofing,\\   Fuzzy\end{tabular} & CNN-LSTM & 682 KB \\ \hline
\cite{thiruloga2022tenet} & 2022 & DL & Unsupervised & Simulation & \textbf{} &  \checkmark & \begin{tabular}[c]{@{}c@{}}Plateau, Continuous\\ Change,Playback,\\  Suppress\end{tabular} & CNN & 59.62 KB / 6064 \\ \hline
\cite{wei2022novel} & 2022 & DL & Unsupervised & OTIDS \cite{lee2017otids} & \textbf{} &  \checkmark  & \begin{tabular}[c]{@{}c@{}}Payload value \\  Manipulation\end{tabular} & AE & N/A \\ \hline
\cite{mansourian2023anomaly} & 2023 & ML / DL & Unsupervised & \begin{tabular}[c]{@{}c@{}}Car Hacking \cite{seo2018gids}, \\  Survival Analysis Dataset   \cite{han2018anomaly}\end{tabular} &  & \checkmark & \begin{tabular}[c]{@{}c@{}}(Gear, RPM) Spoofing, \\  DoS, Fuzzy, \\ Flooding, Malfunction\end{tabular} & LSTM, GNB & N/A \\ \hline
\cite{zhao2022gvids} & 2022 & DL & Unsupervised & Own &  & \checkmark & \begin{tabular}[c]{@{}c@{}}Spoofing, Bus-off, \\  Masquerade, \\ SOME attacks\end{tabular} & GAN & N/A \\ \hline
\multicolumn{10}{c}{\cellcolor[HTML]{dcdcdc}\textbf{CAN Frame-Based Attack Detection}} \\ \hline
\cite{qin2021application} & 2021 & DL & Unsupervised & Own &  \checkmark  &  \checkmark  & \begin{tabular}[c]{@{}c@{}}Random CAN \\  payload Values\end{tabular} & LSTM & N/A\\ \hline
\cite{khan2021enhanced} & 2021 & DL & Unsupervised & Car Hacking \cite{seo2018gids} &  \checkmark  &  \checkmark  & \begin{tabular}[c]{@{}c@{}}DoS, Fuzzy,\\  RPM, Gear \\ Spoofing\end{tabular} & LSTM & N/A  \\ \hline
\cite{kristianto2024sustainable} & 2024 & DL & Unsupervised & Car Hacking \cite{seo2018gids} & \checkmark & \checkmark & \begin{tabular}[c]{@{}c@{}}(Gear, RPM) Spoofing,\\  DoS, Fuzzy\end{tabular} & RNNs and AEs & \begin{tabular}[c]{@{}c@{}}Multiple IDSs (119 -272 ) KB\\  for each gateway\end{tabular} \\ \hline
\cite{cobilean2023anomaly} & 2023 & DL & Self-supervised & \begin{tabular}[c]{@{}c@{}}Survival Analysis \\ Dataset \cite{han2018anomaly}\end{tabular} & \checkmark & \checkmark & Malfunction & Transformer & N/A \\ \hline
\cite{narasimhan2021unsupervised} & 2021 & DL & Unsupervised & ML350 \cite{ML3502019Dataset} & \checkmark & \checkmark & DoS , Fuzzy & AE, GMM & N/A \\ \hline
\cite{wang2021can} & 2021 & ML & Supervised & \begin{tabular}[c]{@{}c@{}}Simulation, \\  OTIDS \cite{lee2017otids}\end{tabular} & \checkmark & \checkmark & (Gear, RPM) Spoofing & FLXGBoost & N/A \\ \hline
\cite{agrawal2022novelads} & 2022 & DL & Unsupervised & Car Hacking\cite{seo2018gids} & \checkmark & \checkmark & \begin{tabular}[c]{@{}c@{}}(Gear, RPM) Spoofing, \\  DoS, Fuzzy\end{tabular} & CNNs, LSTMs & N/A \\ \hline
\cite{Kukkala2020INDRA} & 2020 & DL & Unsupervised & \begin{tabular}[c]{@{}c@{}}SynCAN \\  \cite{hanselmann2020canet}\end{tabular} & \checkmark & \checkmark & \begin{tabular}[c]{@{}c@{}}Flooding, Plateau, Continuous,\\  Suppress, Playback\end{tabular} & GRU AE & 443 kB \\ \hline
\cite{shi2024ids} & 2024 & DL & Unsupervised & \begin{tabular}[c]{@{}c@{}}Car Hacking \cite{seo2018gids}, \\  Car Hacking: Attack \& Defence \\  Challenge 2020  \cite{kang2021car}\end{tabular} & \checkmark & \checkmark & \begin{tabular}[c]{@{}c@{}}(Gear, RPM) Spoofing, \\  DoS, Replay, Fuzzy\end{tabular} & LSTM, CNN, AE & N/A \\ \hline
\cite{longari2020cannolo} & 2020 & DL & Unsupervised & Recan \cite{zago2020recan} & \checkmark & \checkmark & \begin{tabular}[c]{@{}c@{}}Interleave, Discontinuity, \\  Data field anomalies\end{tabular} & LSTM -AE & Less than 10 MB \\ \hline

\cite{longari2023candito} & 2023 & DL & Unsupervised  &  \begin{tabular}[c]{@{}c@{}} Recan \cite{zago2020recan}, \\ car-hacking  
\cite{song2020vehicle} \end{tabular} &  \checkmark & \checkmark  &	 \begin{tabular}[c]{@{}c@{}} (Gear, RPM) Spoofing, \\ DoS, Fuzzy, Masquerade,  \\ Seamless change,\\ Replay \end{tabular} & LSTM AE &  N/A \\ \hline

\cite{hanselmann2020canet} & 2020 & DL & Unsupervised & SynCAN  \cite{hanselmann2020canet} & \checkmark & \checkmark & \begin{tabular}[c]{@{}c@{}}Flooding, Plateau, Continuous,\\  Suppress, Playback\end{tabular} & LSTM AE & N/A \\ \hline
\cite{kishore2022deep} & 2022 & DL & Unsupervised & \begin{tabular}[c]{@{}c@{}}Car Hacking: Attack \& \\ Defence Challenge 2020  \cite{kang2021car} \end{tabular} & \checkmark & \checkmark & \begin{tabular}[c]{@{}c@{}}Flooding, Spoofing,\\   Replay, Fuzzy\end{tabular} & LSTM & N/A \\ \hline
\cite{rajapaksha2023beyond} & 2023 & DL & \begin{tabular}[c]{@{}c@{}}Unsupervised/\\  Supervised\end{tabular} & \begin{tabular}[c]{@{}c@{}}SynCAN  \cite{hanselmann2020canet}, \\ ROAD \cite{verma2020road}\end{tabular} & \checkmark & \checkmark & 13 different attacks & GRU,  Latent AE & 94MB \\ \hline
\cite{kim2023anomaly} & 2023 & DL & Unsupervised & \begin{tabular}[c]{@{}c@{}}Survival Analysis Dataset  \cite{han2018anomaly}, \\ Car Hacking: Attack \& Defence \\  Challenge 2020  \cite{kang2021car}\end{tabular} & \checkmark & \checkmark & \begin{tabular}[c]{@{}c@{}}Spoofing, Replay, \\  Fuzzy\end{tabular} & LSTM-AEs & 3.88 - 3.98 MB \\ \hline
\cite{jo2024intrusion} & 2024 & DL & Unsupervised & Survival Analysis Dataset  \cite{han2018anomaly} & \checkmark & \checkmark & \begin{tabular}[c]{@{}c@{}}Flooding, Fuzzy, \\ Malfunction\end{tabular} & Transformer & N/A \\ \hline
\cite{xiao2019robust} & 2019 & DL & Unsupervised & OTIDS \cite{lee2017otids} & \checkmark & \checkmark & \begin{tabular}[c]{@{}c@{}}DoS, Fuzzy,\\   Impersonation\end{tabular} & ConvLSTM & N/A \\ \hline
\end{tabular} }
\caption{Summary of related work on unknown attack detection methods}
\label{Unknown_attack_detection}
\end{table}

\subsubsection{Limitations of Unknown Attacks Detection}
\label{Limitations_of_Unknown_Attacks_Detection}
All the proposed anomaly detection IDSs are trained on normal data and use binary classification to classify traffic data as either normal or anomalous, detecting any deviations from the normal data. While it is crucial to detect new, previously unknown attacks, as attackers may introduce novel zero-day attacks that do not fit existing patterns, it is equally important to assign fine-grained labels to known attacks. Identifying the specific attack type can be highly beneficial for selecting appropriate countermeasures and conducting post-attack analysis \cite{zhao2022can}. Thus, there is a need for a comprehensive in-vehicle IDS that addresses both known attacks and new, unknown attacks while meeting deployment requirements. To address this, the next section discusses work proposed with the ability to detect and classify known attacks while also identifying new, unknown attacks.

\subsection{Known and Unknown Attacks Detection}
\label{Known_and_Unknown_Attacks_Detection}

\begin{figure}[p]  
\centering
    \rotatebox{90}{\includegraphics[width=\textheight]{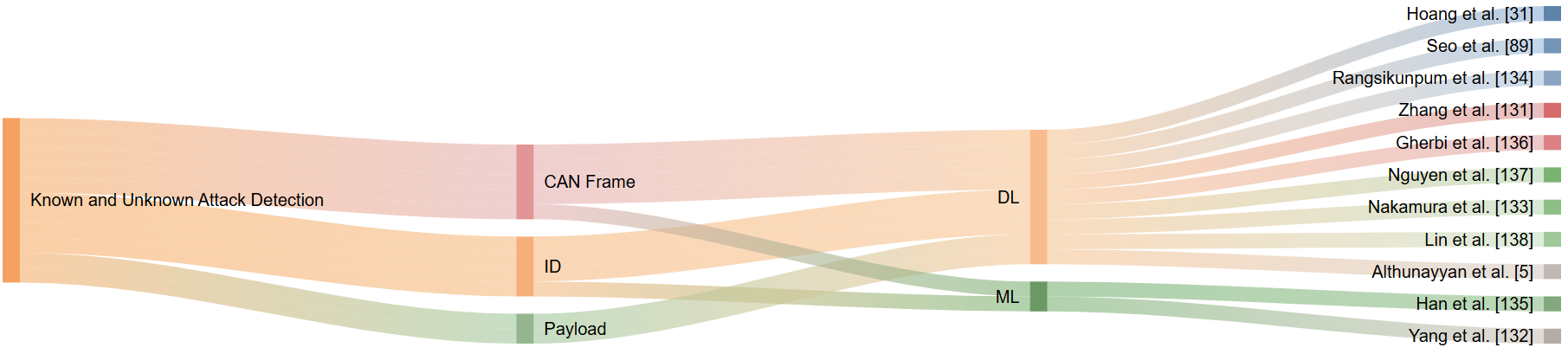}}  
    \caption{Related work on known and unknown attack detection}
    \label{fig:kown_and_unknown_sankey}
\end{figure}

To address the limitations of previous approaches and further improve the robustness and detection capability of in-vehicle IDSs, 10 papers found from the search strategy in Section \ref{Search_Strategy} developed IDSs capable of identifying both known and unknown attacks \cite{zhang2019intrusion, hoang2022detecting, seo2018gids, MTH-IDS2022, nakamura2021vehicle, rangsikunpum2024bids,han2021event, gherbi2020deep,nguyen2024semi,lin2021intrusion}, demonstrating significant advancements in this critical area of cybersecurity. This section reviews state-of-the-art studies and their limitations. Figure \ref{fig:kown_and_unknown_sankey} illustrates exciting work on detecting both known and unknown attacks.

\subsubsection{ID-Based Detection}
This section reviews research where authors used only CAN IDs as the input feature to develop IDSs for detecting both known and new, unknown attacks.

Hoang et al. \cite{hoang2022detecting} and Seo et al. \cite{seo2018gids} have showcased their IDSs’ ability to detect both seen and unseen attacks. However, their IDSs mainly rely on the CAN ID as a singular feature; selecting only the CAN ID feature will limit the detection ability to detect attacks that involve payload manipulation \cite{rajapaksha2023ai}. 

Hoang et al. \cite{hoang2022detecting} propose a lightweight, semi-supervised, learning-based IDS to detect in-vehicle network attacks. The proposed IDS in their study integrates two DL models: autoencoders and generative adversarial networks (GAN). Their IDS was trained on unlabelled data to learn the patterns of normal and malicious data. Only a few labelled samples were used during the subsequent supervised training phase. Even though they use only the CAN ID as the input feature, the number of trainable parameters is 2.15 million for the two models.

Seo et al. \cite{seo2018gids} have developed a GAN-based IDS (GIDS) for in-vehicle network security. The proposed IDS was trained by solely utilising the patterns of CAN IDs from CAN data and then converting the extracted CAN IDs into a simple image. GIDS has two discriminative models to detect both seen and unseen attack data. The first discriminator is specifically trained to handle attacks. In contrast, the second discriminator and the generator are co-trained through an adversarial process. While the generator generates modified images, the second discriminator receives both modified and real CAN images, and its role is to differentiate between the modified and real images.

Rangsikunpum et al. \cite{rangsikunpum2024bids} proposed a Binarized Neural Network (BNN)-based IDS (BIDS) for unknown attack detection and known attack classification, utilizing a BNN and a GAN. The model is hierarchically structured into two stages: attack detection in the first stage and known attack classification in the second. To capture sequential patterns in CAN IDs, consecutive CAN IDs are encoded into one-hot vectors and arranged into a 48 × 48 2D grid. The proposed model is resource-efficient, requiring minimal computational power, making it highly suitable for deployment on low-cost FPGA platforms.

Han et al. \cite{han2021event} proposed an IDS for detecting and identifying abnormalities based on the periodic event-triggered intervals of CAN messages. Statistical features of the event-triggered intervals for each CAN ID, such as mean, variance, quartile deviation, skewness, and kurtosis, were calculated.  These features were then used to train ML models, including DT, RF, and XGBoost to classify attack types. This framework emphasises the event-triggered characteristics of CAN IDs and the statistical moments associated with intervals within a defined time window.

Although using the CAN ID as the only feature reduces the input features and makes the model lightweight, it limits the detection capability of payload manipulation attacks.

\subsubsection{Payload-Based Detection}
This section discusses IDSs that use the 8-byte CAN payload as an input feature to identify both known and new, unknown attacks.

Zhang et al. \cite{zhang2019intrusion} have proposed a DNN-based IDS that aims to automatically extract features for the IDS from the vehicle’s data packets. The authors applied gradient descent with momentum (GDM) and gradient descent with momentum and adaptive gain (GDM/AG) techniques. The study’s results demonstrate the model’s capability to detect replay attacks effectively. The authors accessed the sensor readings, using them as separate features. However, the main limitation of the proposed IDS is that it requires either access to the DBC file or knowledge of the CAN payload, which is confidential and proprietary to the vehicle manufacturer ~\cite{lokman2019intrusion}.

Gherbi et al. \cite{gherbi2020deep} proposed a multivariate time series representation matrix to structure CAN data by integrating flow and payload information. They utilised autoencoder-based DL models such as Fully-Connected Networks (FCNs), CNNs, LSTMs, and Temporal Convolutional Networks (TCNs) to extract hierarchical representation vectors from the CAN matrix for anomaly detection. These vectors are derived either from the bottleneck layer in unsupervised tasks or the final layer in supervised tasks. The findings indicate that TCNs and LSTMs achieve strong performance, demonstrating their ability to effectively capture information from the representation matrix during training.

\subsubsection{CAN Frame-Based Detection}
This section reviews IDSs that use the CAN frame (CAN IDs and payload) as input features to detect both known and new, unknown attacks. 

Nguyen et al. \cite{nguyen2024semi} propose a semi-supervised learning-based IDS that combines a variational autoencoder (VAE) with adversarial environment reinforcement learning (AERL) for multiclass classification. The proposed IDS is able to detect both known and unknown attacks. The objective of this approach is to improve training efficiency by reducing the amount of labelled data required.

Nakamura et al. \cite{nakamura2021vehicle} proposed a hybrid model combining a LightGBM-based supervised model and an autoencoder-based unsupervised model to address the challenge of transferring knowledge across multiple car models for detecting and classifying attacks. Time differences between consecutive CAN IDs, along with CAN ID and payload values, were used as input features. Experimental results showed that the hybrid model outperformed the pre-trained LightGBM model. 

Lin et al. \cite{lin2021intrusion} proposed a two-stage IDS that combines incremental learning (IL) and a DNN, referred to as IL-DNN, to address changes in driving environments and behaviours. In the offline training stage, the DNN was applied to actual CAN data to develop a basic classification model. These predicted class labels were then used in the second stage. In the online detection and updating stage, the DNN model was updated using the IL approach with new, unlabeled data, while simultaneously performing intrusion detection. However, this approach risks degrading model performance if the original model's predictions are incorrect. However, both proposed IDSs in \cite{nakamura2021vehicle} and \cite{lin2021intrusion} are limited to binary classification, do not consider multi-class classification for known attacks, and do not account for the model size.

Most of the aforementioned studies employ either supervised learning-based methods or unsupervised learning-based methods. To leverage the strengths of both approaches, Yang et al. \cite{MTH-IDS2022} have developed a multi-tiered IDS, MTH-IDS, to protect intra-vehicle and external networks from cyberattacks. MTH-IDS use ML algorithms and combines supervised and unsupervised models. The proposed MTH-IDS includes two traditional ML stages: data pre-processing and feature engineering. In the first tier, four tree-based supervised models, DT, RF, ET, and XGBoost, are used to detect known attacks. The second tier incorporates a stacking ensemble model alongside Bayesian optimization using the tree Parzen estimator (BO-TPE) to enhance the accuracy of the base learners. For unknown attack detection, the third tier introduces a novel unsupervised CL-k-means model. Lastly, the fourth tier applies Bayesian optimization with a Gaussian process (BO-GP) and two biased classifiers to refine the performance of the unsupervised learners. Despite achieving good results and a small model size of 2.61 MB, the proposed IDS has certain limitations. In the unsupervised model, the authors add an additional tier with two biased classifiers to improve the results. However, training these biased classifiers on false positives (FPs) and false negatives (FNs) may lead to poorer performance when testing the model on new, unseen data. Furthermore, adding this tier shifts the model from being purely unsupervised, creating a dependency on labelled datasets, which are often challenging to implement in practical scenarios. Moreover, the authors used only four features—CAN ID, and selected three features from the payload field which are DATA[5], DATA[3], and DATA[1] to train the model after feature extraction. Although feature selection approaches may lead to more efficient models, they create the risk that attackers could manipulate features not considered during the model’s training process \cite{kocher2021machine}. This presents a critical limitation in CAN bus data for three reasons. First, selecting a subset of CAN bus payload features as important while discarding others could allow attackers to exploit the neglected features and bypass the model \cite{li2014feature, zhang2015adversarial}. Second, the evolving landscape of attack scenarios means that features chosen to detect one category of attack may become outdated or insufficient to address new, unseen attacks \cite{kocher2021machine}.

Yang et al. \cite{MTH-IDS2022} employed conventional ML models in their proposed IDS due to their lower computational cost compared to DL algorithms. However, DL has shown superior performance in processing large volumes of data efficiently and at a faster rate \cite{jan2019deep}. Considering that modern vehicle ECUs produce around 2,000 CAN frames per second \cite{seo2018gids}, this capability is essential to handle the extensive data of the CAN bus. Moreover, multiple studies have found that DL-based IDSs outperform traditional ML-based IDSs in automotive applications \cite{mehedi2021deep}. This superiority is due to several factors: DL methods are more adaptive, continually being refined with incoming data, which is particularly suitable for the nature of CAN bus data \cite{zhang2019intrusion}. Additionally, traditional ML often requires manual feature engineering, such as applying correlation-based feature selection, which can be time-consuming \cite{nagarajan2023machine}. In contrast, DL automatically deduces features, allowing algorithms to directly discern optimal features from raw data \cite{lampe2023survey}. Furthermore, DL-based IDSs are especially capable of detecting novel attacks and can scale more effectively to highly complex in-vehicle network data while maintaining efficacy \cite{lampe2023survey}. 

To address these limitations, Althunayyan et al. \cite{althunayyan2024robust} proposed a DL-based IDS with a multi-stage approach designed to detect both known and unknown attacks, considering that some attacks may evade detection and be misclassified as normal. The first stage employs a supervised ANN to detect and classify known attacks, while the second stage utilizes an unsupervised LSTM autoencoder to identify unknown attacks that bypass the first model. If the supervised model misclassifies malicious traffic as normal, the anomaly detection model detects deviations from learned patterns and flags them as unseen attacks. Despite incorporating two models, the approach remains lightweight and practical for deployment. Table \ref{known_and_Unknown_attack_detection} summarises the details of known and unknown attack detection studies.

\begin{table*}[!ht]
\centering
\caption{Summary of related work on known and unknown attack detection methods}
\label{known_and_Unknown_attack_detection}
\resizebox{\textwidth}{!}{
\begin{tabular}{ccccccccccc}
\hline
\textbf{Reference} & \textbf{Year} &\textbf{ML / DL} &  \textbf{Category} &  \textbf{Dataset} & \textbf{Algorithm} & \textbf{M-C}&  \textbf{ID} &
\textbf{Payload} & \textbf{\begin{tabular}[c]{@{}c@{}}Model Size/\\  Trainable Parameters\end{tabular}} & \textbf{\begin{tabular}[c]{@{}l@{}}FL \end{tabular}} \\ \hline
\multicolumn{11}{c}{\cellcolor[HTML]{dcdcdc}\textbf{ID-Based Attack Detection}} \\ \hline
\cite{hoang2022detecting} & 2022 & DL&\begin{tabular}[c]{@{}c@{}}Semi-\\supervised \end{tabular} & Car-Hacking \cite{seo2018gids} &\begin{tabular}[c]{@{}c@{}}AE, \\ GAN \end{tabular} &  & \checkmark  &  &2.15 million&  \\ \hline
\\[-.8em]
\cite{seo2018gids} & 2018 & DL & Unsupervised & Car-Hacking \cite{seo2018gids}  & GAN &  & \checkmark  &   &N/A& \\ \hline

\cite{rangsikunpum2024bids} & 2024 & DL & Semi-supervised & \begin{tabular}[c]{@{}c@{}}Car Hacking \cite{seo2018gids}, \\ Survival Analysis Dataset \cite{han2018anomaly}\end{tabular} & BNN, GAN & \multicolumn{1}{c}{\checkmark} & \checkmark &  & 4.07 Mb &  \\ \hline
\cite{han2021event} & 2021 & ML & Unsupervised & Own & DT, RF,  XGBoost & \multicolumn{1}{c}{\checkmark} & \checkmark &  & N/A &  \\ \hline
\multicolumn{11}{c}{\cellcolor[HTML]{dcdcdc}\textbf{Payload-Based Attack Detection}} \\ \hline
\cite{zhang2019intrusion} & 2019 & DL & Supervised & Simulation & DNN &  &   & \checkmark &N/A& \\ \hline

\cite{gherbi2020deep} & 2020 & DL & \begin{tabular}[c]{@{}c@{}}Supervised/\\  Unsupervised\end{tabular} & SynCAN \cite{hanselmann2020canet} & \begin{tabular}[c]{@{}c@{}}FCN, CNN, \\  TCN, LSTM, AE\end{tabular} &  &  & \checkmark & 0.01 - 0.3MB &  \\ \hline
\multicolumn{11}{c}{\cellcolor[HTML]{dcdcdc}\textbf{CAN Frame-Based Attack Detection}} \\ \hline
\cite{nakamura2021vehicle}  & 2021  & DL & Unsupervised & \begin{tabular}[c]{@{}c@{}} Survival Analysis\\ Dataset \cite{han2018anomaly} \end{tabular}  &  \begin{tabular}[c]{@{}c@{}} LightGBM, \\ AE \end{tabular} &  & \checkmark    & \checkmark  & N/A & \\ \hline
\\[-.8em]
\cite{nguyen2024semi} & 2024 & DL & Semi-supervised & \begin{tabular}[c]{@{}c@{}}car-hacking  \cite{song2020vehicle},\\  ROAD \cite{verma2020road}\end{tabular} & VAE and AERL & \multicolumn{1}{c}{\checkmark} & \checkmark & \checkmark & 2,542 KB &  \\ \hline
\cite{lin2021intrusion} & 2021 & DL & \begin{tabular}[c]{@{}c@{}}Supervised/\\ Semi-supervised\end{tabular} & car-hacking  \cite{song2020vehicle} & DNN and IL &  & \checkmark & \checkmark & N/A &  \\ \hline
\cite{MTH-IDS2022}  & 2022 & ML & Hybrid & \begin{tabular}[c]{@{}c@{}}Car-Hacking \cite{seo2018gids},\\ CICIDS2017  \cite{sharafaldin2018toward}\end{tabular}   & \begin{tabular}[c]{@{}c@{}}DT, RF, ET, \\ XGBoost, \\CL-k-means \end{tabular}  & \checkmark &  \checkmark & \checkmark & 2.61 MB & \\ \hline
\cite{althunayyan2024robust}  & 2024 & DL & Hybrid & Car-Hacking \cite{seo2018gids}  & ANN-LSTM AE  & \checkmark &  \checkmark & \checkmark & 2.98 MB / 253,582  &  \checkmark \\ \hline

\end{tabular}}
  \begin{tablenotes}
       \small
       \item \textbf{DL:} Deap Learning, \textbf{FL:} Federated Learning, \textbf{M-C:} Multi-class classification. 
    \end{tablenotes}
\end{table*}

\subsubsection{Limitations of Existing known and Unknown Attacks Detection}
\label{Limitations_of_Existing_known_and_Unknown_Attacks_Detection}
Although we have discussed the limitations of each of the previous work in the previous section, a common limitation of most of the proposed approaches pertains to their deployment strategy. The majority of these studies have implemented their IDSs using a traditional centralised learning approach, which requires transmitting large volumes of data to the cloud for both training and testing on the CAN bus. This method raises significant issues, such as privacy concerns, high communication overhead, and longer response times \cite{chellapandi2023survey}.

\subsection{Evaluation Metics}
\label{Evaluation_Metics}
In this section, we review all the evaluation metrics used to assess the proposed models in previously reviewed papers. The aim is to emphasise the importance of considering these metrics when designing models, rather than focusing on a few while ignoring others, to develop more deployable solutions. Based on the reviewed papers, we categorize the evaluation metrics into performance metrics, time complexity metrics, memory requirement metrics, and other metrics.

Performance metrics assess a model's effectiveness, including accuracy, F1-score, precision, recall (also known as Detection Rate (DR)), Error Rate (ER), confusion matrix, False Negative Rate (FNR), True Positive Rate (TPR), False Positive Rate (FPR), and True Negative Rate (TNR), also known as specificity. Additionally, False Alarm Rate (FAR), Receiver Operating Characteristic (ROC) Curve, Area Under the ROC Curve (AUC-ROC), and Area Under the Precision-Recall Curve (AUPR) are commonly used. These metrics are computed using True Positives (TP), False Positives (FP), False Negatives (FN), and True Negatives (TN). Furthermore, the G-mean score and Matthews Correlation Coefficient (MCC) are valuable for evaluating model performance, particularly in cases of significant class imbalance \cite{guidry2023one, thiruloga2022tenet}. Other relevant metrics include kappa and loss.
For time complexity, several measures are commonly used, including training time, detection (inference) time, and latency. Regarding memory requirement metrics for evaluating model size, key metrics include the number of trainable parameters (which reflects memory usage), the model size in megabytes or kilobytes, and the number of Floating Point Operations (FLOPs).
Other metrics, which are less commonly used in the reviewed papers, include resource allocation, power consumption, and Multiply-Accumulate (MAC) operations, which measure the speed of DL models \cite{le2024multi}.

\begin{table}[t!]
\centering
\resizebox{\textwidth}{!}{
\begin{tabular}{cl }
\hline
\textbf{Metric} & \textbf{Work(s)} \\
\hline
\multicolumn{2}{c}{\textbf{ \cellcolor[HTML]{dcdcdc} Performance Metrics}} \\ \hline

\textbf{Accuracy}   & \begin{tabular}[l]{@{}l@{}} 
 \cite{refat2022detecting}, \cite{sami2020rapid}, \cite{kristianto2024sustainable}, \cite{qin2021application} , \cite{paul2021artificial} ,  \cite{khan2021enhanced} , \cite{MTH-IDS2022},  \cite{nguyen2024semi}, \cite{nakamura2021vehicle} ,  \cite{rangsikunpum2024bids}, \cite{Kukkala2020INDRA}, \cite{mansourian2023anomaly} , \cite{kishore2022deep},  \cite{longari2023candito}, \cite{nguyen2023transformer}, \\ \cite{kalkan2020vehicle}, \cite{kim2023anomaly},  \cite{alalwany2024effective}, \cite{hossain2020long},  \cite{hossain2020effective} , \cite{nandam2022can}, \cite{rangsikunpum2024fpga}, \cite{alalwany2022classification}, \cite{wu2024network}, \cite{fenzl2021vehicle},  \cite{samir2024machine}, \cite{hossain2020lstm}, \cite{hossain2020effective}, \cite{paul2021artificial}, \cite{chougule2024hybridsecnet}, \cite{zhang2024efficient}, \cite{han2021event},  \cite{park2020hierarchical}, \\ \cite{ding2022vehicle}, \cite{shi2024ids}, \cite{seo2018gids}, 
 \cite{alfardus2021intrusion}, \cite{basavaraj2022towards}, \cite{gao2023multi}, \cite{gou2023multi}, \cite{khan2024divacan}, \cite{lin2021intrusion}, \cite{cobilean2023anomaly}, \cite{narasimhan2021unsupervised}, \cite{wang2021can}, \cite{balaji2021neurocan}, \cite{minawi2020machine},   \cite{lin2022using}, \cite{thiruloga2022tenet}, \cite{althunayyan2024robust},  \cite{zhao2022gvids}
  \end{tabular}
 \\ \hline

\textbf{F1-score} & \begin{tabular}[l]{@{}l@{}} \cite{paul2021artificial} , \cite{khan2021enhanced} , \cite{nakamura2021vehicle} , \cite{hossain2020long}, \cite{lin2022using}, \cite{balaji2021neurocan},   \cite{MTH-IDS2022}, \cite{rangsikunpum2024bids}, \cite{jo2024intrusion}, \cite{agrawal2022novelads}, \cite{longari2020cannolo},  \cite{xiao2019robust},   \cite{mansourian2023anomaly} , \cite{kim2023anomaly} ,  \cite{casillo2019embedded}, \\ \cite{khan2024divacan}, \cite{gherbi2020deep}, \cite{nguyen2024semi},  \cite{lin2021intrusion}, \cite{rajapaksha2023beyond},  \cite{hoang2022detecting}, \cite{narasimhan2021unsupervised},  \cite{nazeer2024enhancing}, \cite{sun2021anomaly}, \cite{qin2021application} , \cite{kishore2022deep},  \cite{wang2021can}, \cite{shi2024ids}, 
 \cite{gou2023multi},  \cite{han2021event}, \\ \cite{kalkan2020vehicle}, 
 \cite{alalwany2024effective}, \cite{chougule2024hybridsecnet},  \cite{kishore2024intelligent}, \cite{kristianto2024sustainable}, \cite{song2020vehicle}, \cite{refat2022detecting}, \cite{rangsikunpum2024fpga}, \cite{wu2024network}, \cite{martinelli2017car}, \cite{samir2024machine},  \cite{hossain2020lstm}, \cite{ding2022vehicle},  \cite{hossain2020effective}, \cite{zhang2020convolutional}, \cite{paul2021artificial}, \cite{zhang2024efficient}, \\ \cite{sami2020rapid}, \cite{minawi2020machine}, \cite{alfardus2021intrusion}, \cite{basavaraj2022towards}, \cite{gao2023multi}, \cite{khandelwal2023real},  \cite{park2020hierarchical}, \cite{le2024multi}, \cite{nguyen2023transformer}, \cite{alalwany2022classification}, \cite{nasreldin2021vehicle}, \cite{ma2022gru} , \cite{hossain2020effective} ,  \cite{longari2023candito}, \cite{rajapaksha2022keep}, \cite{cobilean2023anomaly}, \cite{althunayyan2024robust} \end{tabular}
 \\ \hline
\textbf{Precision} &  \begin{tabular}[l]{@{}l@{}} \cite{mansourian2023anomaly}, \cite{khan2021enhanced},  \cite{nakamura2021vehicle} , \cite{nguyen2024semi},  \cite{gherbi2020deep}, \cite{xiao2019robust},  \cite{kishore2022deep},  \cite{rangsikunpum2024bids}, \cite{agrawal2022novelads}, \cite{kristianto2024sustainable},  \cite{qin2021application} , \cite{kim2023anomaly} , \cite{paul2021artificial},\cite{wei2022novel},  \cite{jo2024intrusion}, \\ \cite{hoang2022detecting},  \cite{nguyen2023transformer},  \cite{nazeer2024enhancing},  \cite{casillo2019embedded}, \cite{seo2018gids},  \cite{hossain2020long},  \cite{khan2024divacan}, \cite{ma2022gru} , \cite{gou2023multi}, \cite{ding2022vehicle}, \cite{alalwany2022classification}, \cite{chougule2024hybridsecnet},  \cite{zhang2020convolutional},  \cite{kalkan2020vehicle}, \cite{park2020hierarchical},  \cite{le2024multi}, \cite{song2020vehicle}, \cite{longari2020cannolo}, \\ \cite{kishore2024intelligent},  \cite{refat2022detecting}, \cite{rangsikunpum2024fpga},  \cite{wu2024network},  \cite{martinelli2017car}, \cite{narasimhan2021unsupervised} ,  \cite{fenzl2021vehicle}, \cite{samir2024machine},  \cite{hossain2020lstm}, \cite{hossain2020effective}, \cite{paul2021artificial}, \cite{sami2020rapid}, \cite{sun2021anomaly}, \cite{zhao2022gvids}, \cite{nasreldin2021vehicle}, \cite{gao2023multi},  \cite{alalwany2024effective}, \\ \cite{longari2023candito}, \cite{shi2024ids}, \cite{basavaraj2022towards}, \cite{cobilean2023anomaly} , \cite{khandelwal2023real},  \cite{lin2022using},  \cite{wang2021can}, \cite{althunayyan2024robust} \end{tabular}
\\ \hline

\textbf{Recall} &  \begin{tabular}[l]{@{}l@{}}  \cite{paul2021artificial} ,  \cite{khan2021enhanced},   \cite{nakamura2021vehicle} ,  \cite{gherbi2020deep}, \cite{nguyen2024semi}, \cite{xiao2019robust},  \cite{kishore2022deep},  \cite{qin2021application} , \cite{mansourian2023anomaly} , \cite{shi2024ids},  \cite{nguyen2023transformer}, \cite{rangsikunpum2024bids},  \cite{jo2024intrusion}, \cite{nazeer2024enhancing}, \\ \cite{khandelwal2023real},  \cite{zhao2022gvids},    \cite{wang2021can},  \cite{casillo2019embedded}, \cite{lin2022using},  \cite{khan2024divacan},  \cite{longari2020cannolo}, \cite{hoang2022detecting},  \cite{ma2022gru}, \cite{gou2023multi}, \cite{kalkan2020vehicle},  \cite{chougule2024hybridsecnet}, \cite{kishore2024intelligent},  \cite{alalwany2024effective},  \cite{ding2022vehicle},  \cite{zhang2020convolutional}, \\ \cite{park2020hierarchical},  \cite{song2020vehicle}, \cite{refat2022detecting}, \cite{rangsikunpum2024fpga},  \cite{alalwany2022classification}, \cite{wu2024network},  \cite{martinelli2017car}, \cite{hossain2020lstm}, \cite{agrawal2022novelads},  \cite{gao2023multi}, \cite{paul2021artificial},\cite{sami2020rapid}, \cite{le2024multi}, \cite{basavaraj2022towards}, \cite{nasreldin2021vehicle}, \cite{minawi2020machine}, \cite{hossain2020long},  \\ \cite{kristianto2024sustainable}, \cite{kim2023anomaly}, \cite{sun2021anomaly}, \cite{cobilean2023anomaly}, \cite{narasimhan2021unsupervised} 
, \cite{hossain2020effective} ,  \cite{MTH-IDS2022} ,\cite{longari2023candito},  \cite{seo2018gids}, \cite{althunayyan2024robust} \end{tabular}\\ \hline

\textbf{Confusion Matrix} & \begin{tabular}[l]{@{}l@{}} \cite{khandelwal2023real},  \cite{avatefipour2019intelligent},  \cite{nakamura2021vehicle} , \cite{kishore2024intelligent},\cite{nguyen2023transformer},  \cite{aksu2022mga}, \cite{song2020vehicle}, \cite{gao2023multi},  \cite{rangsikunpum2024fpga}, \cite{hossain2020lstm}, \cite{samir2024machine}, \\ \cite{le2024multi}, \cite{alalwany2022classification}, \cite{park2020hierarchical}, \cite{casillo2019embedded}, \cite{nazeer2024enhancing}, \cite{hossain2020long}, \cite{gou2023multi}, \cite{khan2024divacan},  \cite{rangsikunpum2024bids}, \cite{MTH-IDS2022}, \cite{zhang2019intrusion}, \cite{althunayyan2024robust} \end{tabular}\\ \hline

\textbf{TPR} & \cite{hossain2020lstm}, \cite{fenzl2021vehicle}, \cite{zhang2024efficient}, \cite{longari2023candito}, \cite{paul2021artificial} ,  \cite{hanselmann2020canet}, \cite{sami2020rapid}, \cite{alfardus2021intrusion}, \cite{rajapaksha2023beyond}, \cite{zhang2019intrusion} 
 \\ \hline
\textbf{FPR} &  \begin{tabular}[l]{@{}l@{}} \cite{samir2024machine}, \cite{Kukkala2020INDRA}, \cite{rajapaksha2022keep}, \cite{nguyen2024semi}, \cite{zhang2024efficient},  \cite{paul2021artificial} , \cite{hossain2020effective},  \cite{sami2020rapid}, \cite{martinelli2017car}, \cite{wu2024network}, \cite{hossain2020lstm}, \cite{hossain2020effective}, \\ \cite{hossain2020long}, \cite{minawi2020machine}, \cite{alfardus2021intrusion}, \cite{longari2023candito}, \cite{khandelwal2023real}, \cite{shi2024ids}, \cite{zhao2022gvids} , \cite{kishore2024intelligent}, \cite{rajapaksha2023beyond}, \cite{zhang2019intrusion}, \cite{kang2016intrusion} \end{tabular}  \\ \hline

\textbf{FNR} & \cite{hossain2020effective} , \cite{khandelwal2023real},  \cite{paul2021artificial} , \cite{shi2024ids}, \cite{song2020vehicle}, \cite{samir2024machine}, \cite{wu2024network}, \cite{nguyen2024semi}, \cite{kang2016intrusion}, \cite{hossain2020effective}, \cite{hossain2020long}, \cite{zhang2020convolutional}, \cite{thiruloga2022tenet}, \cite{le2024multi}, \cite{rajapaksha2023beyond}, \cite{rajapaksha2022keep}, \cite{ding2022vehicle}, \cite{zhao2022gvids}
 \\ \hline

\textbf{TNR} &  \cite{samir2024machine}, \cite{hanselmann2020canet}, \cite{paul2021artificial} , \cite{rajapaksha2023beyond}, \cite{nazeer2024enhancing}  \\ \hline

\textbf{ER} & \cite{song2020vehicle}, \cite{ding2022vehicle}, \cite{hoang2022detecting}, \cite{sun2021anomaly}, \cite{nguyen2023transformer}
\\ \hline
\textbf{ROC} & \cite{kang2016intrusion}, \cite{balaji2021neurocan}, \cite{avatefipour2019intelligent},  \cite{xiao2019robust},  \cite{jo2024intrusion}, \cite{chougule2024hybridsecnet}, \cite{alalwany2022classification}, \cite{qin2021application} , \cite{nguyen2023transformer}, \cite{martinelli2017car}, \cite{aksu2022mga}, \cite{zhang2019intrusion} \\ \hline

\textbf{AUC-ROC} & \begin{tabular}[l]{@{}l@{}}\cite{paul2021artificial} ,\cite{wei2022novel}, \cite{kishore2024intelligent}, \cite{kim2023anomaly} , \cite{refat2022detecting}, \cite{hossain2020lstm}, \cite{samir2024machine}, \cite{ding2022vehicle}, \cite{khan2021enhanced}, \cite{alalwany2024effective}, \cite{hossain2020long},  \cite{sami2020rapid},  \\ \cite{boumiza2019anomaly} , \cite{thiruloga2022tenet}, \cite{lin2022using}, \cite{hanselmann2020canet}, \cite{kishore2022deep},  \cite{han2021event}, \cite{nazeer2024enhancing}, \cite{seo2018gids},  \cite{wang2021can}, \cite{longari2020cannolo} \end{tabular}  \\ \hline

\textbf{AUPR} & \cite{wei2022novel} \\ \hline
\textbf{MCC} & \cite{thiruloga2022tenet, longari2023candito} \\ \hline
\textbf{TP, FP} &  \cite{nandam2022can}, \cite{avatefipour2019intelligent} \\ \hline
\textbf{FN, TN} & \cite{avatefipour2019intelligent} \\ \hline
\textbf{G-mean Score} & \cite{guidry2023one} \\ \hline
\textbf{FAR} & \cite{MTH-IDS2022}, \cite{althunayyan2024robust} \\ \hline
\textbf{Kappa} & \cite{khan2021enhanced} \\ \hline
\textbf{Loss} & \cite{chougule2024hybridsecnet}, \cite{basavaraj2022towards} \\ \hline
\multicolumn{2}{c}{\textbf{ \cellcolor[HTML]{dcdcdc} Time Complexity Metrics}} \\ \hline
\textbf{Training Time} & \cite{song2020vehicle}, \cite{park2020hierarchical}, \cite{basavaraj2022towards}, \cite{khan2021enhanced} , \cite{lin2022using}, \cite{nguyen2023transformer}, \cite{avatefipour2019intelligent}, \cite{kristianto2024sustainable}, \cite{zhang2019intrusion}, \cite{gherbi2020deep}, \cite{kang2016intrusion}, \cite{MTH-IDS2022} \\
\hline
\textbf{\begin{tabular}[c]{@{}l@{}}Detection Latency/ \\Inference Time  \end{tabular}} & \begin{tabular}[l]{@{}l@{}} \cite{song2020vehicle}, \cite{rangsikunpum2024fpga}, \cite{kang2016intrusion}, \cite{le2024multi}, \cite{rajapaksha2023beyond}, \cite{zhang2024efficient}, \cite{park2020hierarchical}, \cite{chougule2024hybridsecnet}, \cite{nguyen2023transformer}, \cite{avatefipour2019intelligent}, \cite{longari2023candito}, \cite{rajapaksha2022keep}, \cite{khandelwal2023real},  \cite{khan2021enhanced} , 
\cite{balaji2021neurocan}, \cite{sun2021anomaly}, 
\cite{thiruloga2022tenet}, \\ \cite{zhao2022gvids}, \cite{kristianto2024sustainable},  \cite{agrawal2022novelads}, \cite{Kukkala2020INDRA}, \cite{shi2024ids}, \cite{kim2023anomaly}, \cite{jo2024intrusion}, \cite{hoang2022detecting}, \cite{seo2018gids}, \cite{rangsikunpum2024bids}, \cite{zhang2019intrusion}, \cite{nguyen2024semi}, \cite{ma2022gru} ,\cite{lin2021intrusion}, \cite{MTH-IDS2022}  \end{tabular}\\
\hline
\textbf{Execution Time} & \cite{fenzl2021vehicle}, \cite{aksu2022mga}, \cite{minawi2020machine}, \cite{alfardus2021intrusion}, \cite{gou2023multi}, \cite{khan2024divacan}, \cite{han2021event}, \cite{zhang2019intrusion} \\
\hline
\multicolumn{2}{c}{\textbf{\cellcolor[HTML]{dcdcdc} Memory Requirements Metrics}} \\ \hline
\textbf{Memory Footprint (Size)} & \cite{rangsikunpum2024fpga}, \cite{zhang2024efficient}, \cite{sun2021anomaly}, \cite{thiruloga2022tenet}, \cite{kristianto2024sustainable}, \cite{Kukkala2020INDRA}, \cite{longari2020cannolo}, \cite{rajapaksha2023beyond}, \cite{kim2023anomaly}, \cite{rangsikunpum2024bids}, \cite{gherbi2020deep}, \cite{khan2021enhanced} , \cite{nguyen2024semi}, \cite{MTH-IDS2022}, \cite{althunayyan2024robust}\\ \hline
\textbf{Parameters} & \cite{le2024multi}, \cite{thiruloga2022tenet}, \cite{kristianto2024sustainable}, \cite{rajapaksha2023beyond}, \cite{hoang2022detecting}, \cite{gherbi2020deep}, \cite{althunayyan2024robust} \\ \hline
\textbf{Training Cost (FLOPS)} & \cite{nguyen2023transformer}, \cite{kristianto2024sustainable} \\ \hline
\multicolumn{2}{c}{\textbf{ \cellcolor[HTML]{dcdcdc} Other Metrics}} \\ \hline
\textbf{Resource Utilization} & \cite{shi2024ids}, \cite{Kukkala2020INDRA}, \cite{khandelwal2023real}, \cite{rangsikunpum2024fpga} \\ \hline
\textbf{\begin{tabular}[c]{@{}l@{}}Power/Energy \\ Consumption \end{tabular}} & \cite{kristianto2024sustainable}, \cite{khandelwal2023real}, \cite{rangsikunpum2024fpga} \\ \hline
\textbf{MAC} & \cite{le2024multi} \\ \hline

\end{tabular}}
\caption{Evaluation metrics used in existing works}
\label{Evaluation_metrics}
\end{table}

Table \ref{Evaluation_metrics} shows each evaluation metric used in the reviewed papers. Most studies have focused on some performance metrics while giving less consideration to time and memory requirements. Considering all these metrics (performance, time, and memory) makes the proposed models more deployable and easier to compare with other works.

\section{Federated Learning for In-Vehicle Networks}
\label{Federated_Learning_Section}
This section starts with an overview of the FL approach, followed by a review of existing FL-based in-vehicle IDSs, and concludes with their limitations.

\subsection{Overview of Federated Learning}
FL is a privacy-preserving decentralised learning technique that trains models locally without transferring row data to a centralised server \cite{li2020review}. Instead, it transfers model parameters to a centralised server, which aggregates the clients' models to build a shared global model \cite{agrawal2022federated}. This integration of FL into IDSs enhances security and privacy, addressing the growing challenges of protecting data in an increasingly interconnected world. While ML and DL have made notable progress in in-vehicle IDSs, it is crucial to recognise their limitations, particularly regarding data privacy and communication efficiency. FL mitigates these challenges by enabling local model training while preserving the privacy of raw data \cite{alsamiri2023federated}. FL is well-suited for in-vehicle IDSs for several compelling reasons:
\begin{itemize}
    \item The FL approach preserves data privacy by periodically transmitting learned model parameters to the cloud server instead of sharing raw data. This aligns with various data protection regulations, such as GDPR (Europe), CCPA (California), PIPEDA (Canada), and LGPD (Brazil), which are designed to prevent the unauthorised transfer of sensitive information.

    \item FL allows multiple participants to efficiently develop a robust global model while preserving user data privacy. It enables real-time model updates and data access without the need to communicate with a central server.

    \item FL reduces latency by avoiding sending raw data to a central server \cite{agrawal2022federated}.

    \item Referring to the 2020 guidelines of the International Telecommunication Union \cite{x1375Guidelines} for IDS in vehicular networks, an in-vehicle IDS must have the ability to regularly update its set of rules.

    \item FL improves the adaptability of IDS to new, previously unseen attacks by incorporating local models updated with those trained on newly detected attacks. This enables the continuous updating of models as new data becomes available, ensuring effective response to evolving threats in real-time.
    
    \item FL enables the development of a universal model that covers diverse driving scenarios, vehicle states, and driving behaviors \cite{althunayyan2024hierarchical}.  
\end{itemize}

\begin{figure*}[t!]
\centering
\includegraphics[scale=.13]{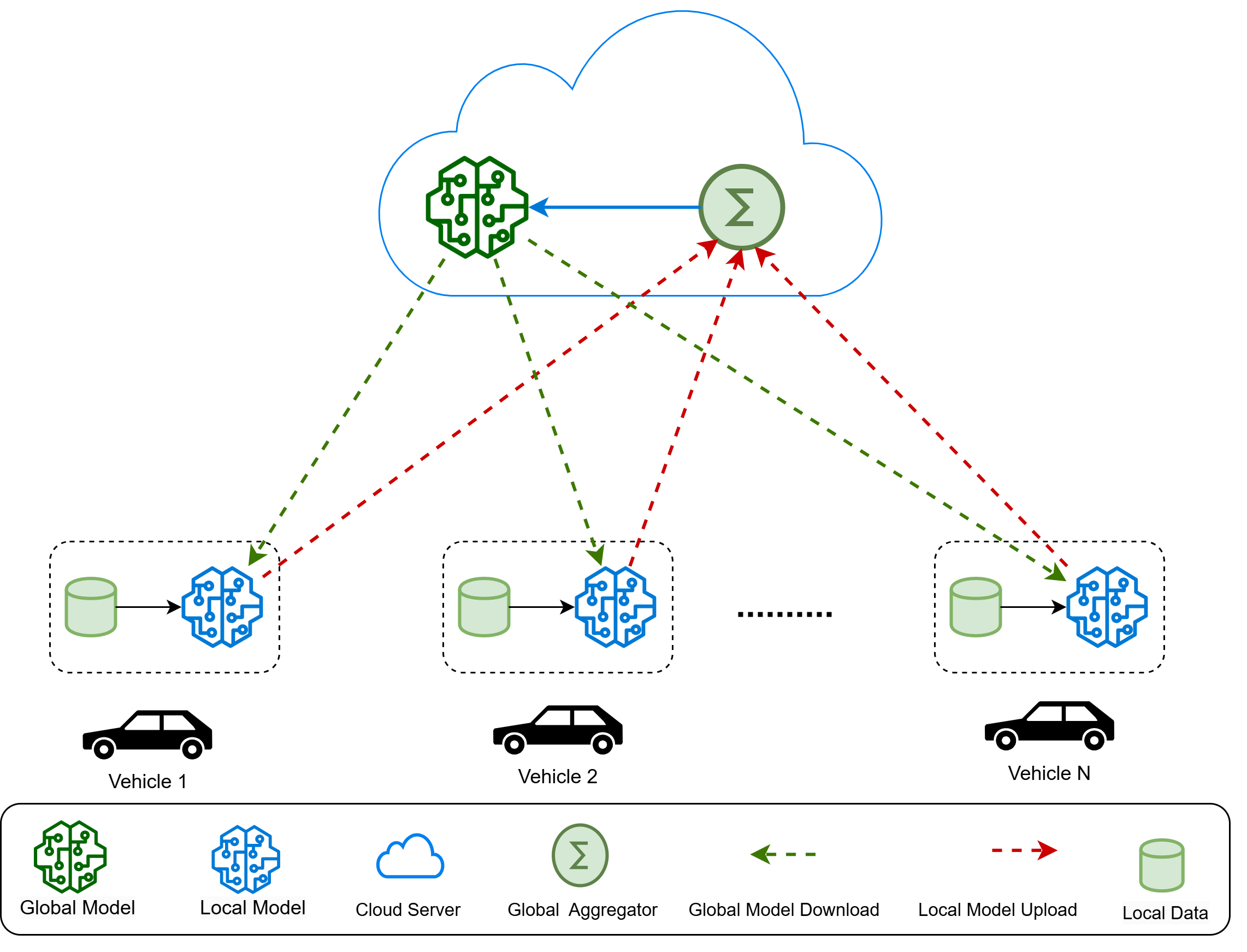}
\caption{Federated Learning Architecture}
\label{Federated_Learning_Framework}
\end{figure*}

As depicted in Figure ~\ref{Federated_Learning_Framework}, the standard cloud-based FL architecture consists of a cloud server and multiple \textit{N} clients (vehicles). Selected clients download the global model from the server, perform several rounds of local training using their own private data, and subsequently return the updated model weights to the server for aggregation.  This iterative process continues until the model reaches the desired level of accuracy.

\subsection{Federated Learning for Intrusion Detection Systems for In-Vehicle Networks} 
\label{Related_Work_on_Federated_Learning_IDSs}

Driss et al. \cite{driss2022federated} introduced an FL-based framework for detecting attacks in vehicular sensor networks. The authors highlighted the importance of lightweight security solutions, recognising the resource limitations of smart sensing devices in these networks. To tackle this challenge, they employed a combination of Gated Recurrent Units (GRU) and an ensemble method using RF to aggregate the global ML models. The dataset was evenly distributed among the clients.

Shibly et al. \cite{shibly2022personalized} proposed a personalised FL-based IDS that eliminates the need for data sharing. The authors explored both supervised and unsupervised methods within the FL framework, including CNN, XGBoost, MLP, and AE. Although their results were promising for both binary and multiclass classification, they did not account for non-IID data distributions.

Yu et al. \cite{yu2022federated} presented an FL-based IDS using LSTM for in-vehicle networks. They leveraged the periodicity of CAN communications to forecast the arbitration IDs of incoming messages. The 11-bit arbitration ID is converted into vectors through one-hot encoding, which are then used by the LSTM to predict the next arbitration ID. The data is equally divided among clients, with each client containing 1,000 instances for training and 200 for testing. A comparison between the FL-based and centralised IDS showed a 0.071 accuracy reduction for the FL-based IDS. However, the authors proposed that this reduction could be addressed with a cumulative error scheme.

Zhang et al. \cite{zhang2023federated} designed an anomaly detection system using a graph neural network, able to detect CAN bus intrusions in just 3 milliseconds. The IDS utilises a two-stage classifier cascade, with one classifier dedicated to anomaly detection within a single class and the other to classifying attacks into multiple categories. An openmax layer is incorporated into the multi-class classifier to handle novel anomalies from unseen classes.

Yang et al. \cite{yang2022federated} developed an IDS for in-vehicle networks using federated deep learning. Their approach capitalises on the periodicity of network messages, incorporates the ConvLSTM model, and trains the model via federated DL. To simulate a non-IID environment, clients were given different numbers of data samples (ranging from 50 to 3500), though details on how the data was distributed among clients and across classes were not specified.

Taslimasa et al. \cite{taslimasa2023imagefed} introduced ImageFed, a privacy-preserving IDS that employs federated CNNs. To create a non-IID environment, data were allocated to vehicles using a Dirichlet$(\mu)$ distribution, with $(\mu)$ values varying from 0.1 to 0.7. To assess ImageFed's resilience, they investigated two potential scenarios that could cause a decline in FL performance: non-IID clients and restricted access to training data.

Longari et al. \cite{longari2023candito} deployed their proposed IDS, CANdito, presented in Section \ref{Unknown_Attacks_Detection}, in an FL setting to evaluate its detection efficiency and communication overhead, comparing it to a centralised version of the same algorithm. Experimental results suggest that FL could be a suitable approach in real-world scenarios where data privacy and security cannot be ignored. While the detection capabilities of the federated model are slightly lower than those of the centralised model, it still demonstrates robust performance. 

To overcome the challenge of DL models requiring large amounts of data to achieve optimal performance—particularly in the case of CAN bus IDS—Hoang et al. \cite{hoang2023canperfl} proposed CANPerFL, an IDS that employs a personalised FL approach to aggregate datasets from different car models. Their approach builds a universal model trained on a small amount of data from each manufacturer, providing global knowledge that enhances the performance of individual participants. Experimental results show that the proposed model improves F1 scores by 4\% overall compared to baselines. Moreover, it offers significant advantages when the local dataset of each participant is relatively small.

Althunayyan et al. \cite{althunayyan2024hierarchical} deployed their proposed IDS from \cite{althunayyan2024robust} within a Hierarchical FL (H-FL) framework. This framework aims to address the limitations of standard FL-based IDSs, which rely on a single central aggregator, leading to performance bottlenecks and introducing a single point of failure that compromises robustness and scalability. By incorporating multiple edge aggregators along with the central aggregator, the proposed H-FL mitigates the risk of single-point failures, enhances scalability, and optimises the distribution of computational load. Experimental results demonstrate that deploying the IDS within the H-FL framework can improve the F1-score by up to 10.63\%, effectively overcoming the limitations of edge-FL in terms of dataset diversity and attack coverage.

Table \ref{tab:FL_Related_Wroks} summarises previous work. In cases where the aggregation function is not explicitly stated, as in \cite{driss2022federated, shibly2022personalized}, it is assumed that FedAvg was employed.

\begin{table}[!ht]
    \centering
    \resizebox{\textwidth}{!}{%
        \begin{tabular}{cccccc}
            \hline
            \textbf{Reference} & \textbf{FL} &\textbf{ Non-IID} & \textbf{\begin{tabular}[c]{@{}c@{}} Aggregation \\Function \end{tabular}} & \textbf{Dataset} & \textbf{FL Implementation}\\ \hline
             \cite{taslimasa2023imagefed} & Standard & \checkmark & FedAvg & car-hacking  \cite{song2020vehicle} & PyTorch\\  \hline

            \cite{yu2022federated} & Standard & x & FedAvg & HCRL CAN Intrusion Detection \cite{lee2017otids} & N/A\\  \hline

            \cite{shibly2022personalized} & Standard & x & FedAvg & \begin{tabular}[c]{@{}c@{}} car-hacking \cite{song2020vehicle},\\ NAIST CAN attack dataset\cite{hossain2020lstm} \end{tabular} & Keras, TensorFlow\\  \hline

            \cite{zhang2023federated} & Standard  & N/A & FedAvg, FedProx & READ \cite{marchetti2018read} & N/A\\  \hline
            
            \cite{driss2022federated} & Standard & x & FedAvg & \begin{tabular}[c]{@{}c@{}}Car Hacking: Attack \& Defence\\ Challenge 2020  \cite{kang2021car}\end{tabular} & Keras, TensorFlow\\  \hline
           
            \cite{yang2022federated} & Standard & \checkmark & FedAvg & HCRL CAN Intrusion Detection \cite{lee2017otids} & N/A \\  \hline

             \cite{longari2023candito} & Standard & N/A  & FedAvg, FedProx & Recan \cite{zago2020recan} & N/A \\  \hline
             
            \cite{hoang2023canperfl} & Standard & -  & FedAvg & Own & Pytorch, Flower \\  \hline

             \cite{althunayyan2024hierarchical}& Hierarchical  & \checkmark  & FedAvg &\begin{tabular}[c]{@{}c@{}} car-hacking ~\cite{song2020vehicle},\\ Car Hacking \cite{seo2018gids} \end{tabular} & Flower\\  \hline
        
        \end{tabular}%
    }
     \caption{FL-based IDSs for in-vehicle network}
    \label{tab:FL_Related_Wroks}
\end{table}

\subsection{Limitations of Existing FL-Based IDSs}
\label{Limitations_of_Existing_FL-Based_IDSs}
Although previous works have contributed to the field of FL-based in-vehicle IDSs, they exhibit certain limitations. A major challenge in FL is managing non-independent and identically distributed (Non-IID) data, where the training data on each client varies significantly, leading to differing data distributions among clients \cite{mcmahan2017communication}. In real-world applications, data is typically Non-IID due to variations in user behaviour, preferences, and environments \cite{li2022federated}. However, most existing studies do not account for Non-IID data and instead assume data partitions where clients receive either an equal number of samples or samples from all classes (i.e., types of attacks). This assumption contradicts real-world FL scenarios, which inherently involve Non-IID data distributions \cite{hernandez2023intrusion}, resulting in an unrealistic evaluation of FL-based IDS performance \cite{zhao2018federated}.
Only a few studies \cite{yang2022federated, taslimasa2023imagefed, althunayyan2024hierarchical} have explicitly considered Non-IID data distributions. In \cite{yang2022federated}, nine candidate clients are assumed, each possessing varying numbers of data samples (50, 100, 150, 1000, 1500, 2000, 2500, 3000, 3500), but the distribution of samples across classes and among clients remains unclear. In contrast, \cite{taslimasa2023imagefed} and \cite{althunayyan2024hierarchical} implement a Non-IID setting by distributing data to vehicles using a \textit{Dirichlet}$(\mu)$ distribution, where the $(\mu)$ parameter is adjusted between 0.1 and 0.7 to control the level of Non-IIDness. Another key limitation in FL-based in-vehicle IDS research is the lack of client selection strategies. Real-world FL scenarios involve clients with varying resources, network stability, and data quality. However, existing studies assume equal participation in every training round, ignoring the dynamic nature of vehicular environments and the need for adaptive selection.

\section{Future Research Directions}
\label{Future_Research_Directions}
Based on the survey in the previous sections, this section identifies the limitations of existing approaches and explores potential future research directions for enhancing the security of in-vehicle networks.

\begin{itemize}
    \item \textbf{Limited Access to Real-World Datasets:} It is a fact that the best ML/DL-based models are derived from high-quality data. Therefore, a key challenge in in-vehicle security research is the limited access to real-world datasets that reflect diverse driving behaviours and environments, such as urban, mountainous, and rural terrains. Existing datasets fail to capture the full complexity of real-world driving conditions, primarily due to privacy and legal constraints \cite{said2020network}. Consequently, most proposed IDSs have been trained and evaluated under restricted conditions, limiting their ability to generalise normal vehicle behaviour across varied scenarios. Moreover, the literature review highlights that publicly available datasets are often less challenging, allowing even simple ML models to achieve high accuracy. However, the effectiveness of these ML/DL-based IDSs in real-world applications may not be guaranteed. Since in-vehicle networks demand high reliability, this could hinder their practical implementation. A promising research direction is the exploration of streaming learning, which enables models to dynamically adapt in real-time as vehicles encounter different driving conditions. This approach could enhance detection accuracy and improve system adaptability across diverse environments.

     \item \textbf{Protecting the in-vehicle IDSs:} In-vehicle IDSs are vulnerable to adversarial attacks, as recent studies \cite{aloraini2024adversarial, li2021adversarial} have highlighted the vulnerabilities of these systems. Adversarial attacks manipulate input data to deceive models into producing incorrect or misclassified outputs \cite{li2021adversarial}, posing significant risks to the safety and security of CAVs. From the literature, it is evident that almost no proposed IDS has considered protecting the system from adversarial attacks, except for the work in \cite{li2021adversarial}, where Li et al. \cite{li2021adversarial} developed a defense strategy to protect LSTM-based IDSs from adversarial attacks. Consequently, deploying IDSs without properly evaluating their adversarial robustness not only fails to protect the vehicle but also potentially escalates the risk of vehicle manipulation. Thus, training IDSs on adversarial samples to detect these attacks is a possible solution. Moreover, adapting defense strategies from other fields could significantly enhance the resilience of in-vehicle IDSs, ensuring robustness against both known and emerging threats, including adversarial examples. This remains a crucial area for future research.

     \item \textbf {False Positives in Unsupervised Learning:} As with all unsupervised learning methods \cite{rajapaksha2023ai}, anomaly detection models usually suffer from false positives. In critical systems, minimising false alarms is essential for maintaining system reliability. Some existing approaches train biased classifiers to reduce false positives and false negatives, but this shifts the model away from being purely unsupervised. Future research should focus on finding practical solutions that reduce false positives without compromising the model’s unsupervised nature. One potential direction is to leverage eXplainable AI (XAI) techniques to make the behaviour of in-vehicle IDSs more interpretable and transparent. While AI methods have shown great potential in combating cyberattacks, they often generate false alarms and produce decisions that are difficult to interpret, leading to uncertainty and distrust \cite{axelsson2006understanding}. XAI methods, such as SHapley Additive exPlanations (SHAP) or Local Interpretable Model-agnostic Explanations (LIME), can provide clearer insights into the decision-making process of IDSs, allowing for better responses to alarms and fostering greater trust in AI-driven security systems \cite{lundberg2022experimental}. Further exploration of XAI could significantly improve both the transparency and reliability of AI-based in-vehicle IDSs.

    \item \textbf{Vehicle-Specific Models and Generalisation Challenges:} Another limitation is the assumption that all vehicles in the FL environment share the same make, model, CAN IDs, and payload interpretations. This assumption could necessitate developing separate models for each vehicle make and model, leading to increased complexity. Generalising the IDS to learn across different vehicle types, rather than relying on distinct models for each, remains a significant challenge due to variations in CAN bus data and the lack of access to DBC files, which define signal meanings. While FL has shown promise in enhancing IDS performance by integrating models from diverse driving scenarios and vehicle states, achieving robust model generalisation across all vehicle types is complex. Future research could explore techniques such as domain adaptation or transfer learning to bridge the gap between different vehicle models and make the system more general across all vehicle types.

     \item \textbf {Client Selection in FL:} Another future direction for improving the efficiency of the FL process is exploring methods for selecting or excluding clients. Given the heterogeneity of in-vehicle network traffic, it is neither practical nor efficient to include all vehicles as federated clients \cite{yang2022federated}. Investigating effective client selection strategies is crucial to optimising model accuracy while minimising computational and communication overhead. Based on the reviewed papers on FL-based in-vehicle IDS, no work has been done on client selection using in-vehicle traffic data. Potential strategies could involve selecting clients based on similarities in CAN bus data, driving behaviour, or geographical area to ensure that the FL process remains efficient.

    \item \textbf { Evaluation Metrics:} The majority of the reviewed literature focused on evaluating their proposed IDSs using performance metrics such as accuracy, F1-score, precision, and recall. However, many existing IDSs either fail to consider memory constraints and real-time requirements when designing in-vehicle IDSs \cite{rajapaksha2023ai}, making many proposed IDSs impractical for real-world applications. Given the memory constraints of ECUs and the real-time requirements in in-vehicle networks \cite{kristianto2024sustainable}, an efficient IDS must be lightweight, have a small memory footprint \cite{zhang2024efficient, Kukkala2020INDRA}, and satisfy real-time performance requirements. When designing in-vehicle IDS solutions, it is essential to consider the deployment requirements \cite{lokman2019intrusion}. The development and deployment of IDSs are significantly impacted by the constraints of ECUs in in-vehicle networks, which include limited memory storage, computing power, and bandwidth \cite{rajapaksha2023ai}. Moreover, since CAN is a time-critical system, inference time and detection latency are essential safety-related metrics for in-vehicle IDS to ensure real-time performance. Inference time refers to the amount of time required for a trained model to generate predictions on a new data batch \cite{le2024multi}. Latency, on the other hand, is the time taken for a packet to travel from its source to its destination \cite{MTH-IDS2022}. The United States (US) Department of Transportation states that critical vehicle safety services, such as collision and attack warnings, should operate with a latency of 10 to 100 ms \cite{abualhoul2016visible}. Meanwhile, Vehicle-to-everything (V2X)-based autonomous and cooperative driving applications require even stricter latency, typically between 10 and 20 ms \cite{moubayed2020edge}. Thus, for a vehicle-level IDS, the time required to process each network packet must be less than 10 ms to meet real-time requirements.

\end{itemize}

\section{Conclusion}
\label{Conclusion}
CAVs improve transportation efficiency but are vulnerable to cybersecurity threats, particularly due to the insecurity of the CAN bus protocol. These cyberattacks can have severe consequences, such as compromising control over essential systems, necessitating robust and reliable security measures. ML-based in-vehicle IDSs offer an effective solution by detecting malicious activities in real time.

The main contribution of this paper is a comprehensive survey of existing ML and DL approaches for building in-vehicle IDSs, focusing on detecting known attacks (38 papers), unknown attacks (27 papers), and combined known and unknown attacks (11 papers). Moreover, we reviewed the evaluation metrics used by researchers to build their IDSs and categorised them into performance metrics, time complexity metrics, memory requirement metrics, and other metrics, emphasizing the importance of considering all these metrics to achieve more deployable solutions.

Additionally, we reviewed research on FL-based IDSs (9 papers) applied to in-vehicle networks. The total number of reviewed papers in this survey is 85. Lastly, we present future directions that can help enhance the security and privacy of in-vehicle IDSs.

\section*{Declaration of Generative AI and AI-assisted technologies in the writing process}
During the preparation of this work, the authors used ChatGPT-4 in order to improve readability and language. After using this tool/service, the authors reviewed and edited the content as needed and take full responsibility for the content of the publication.

\bibliographystyle{unsrtnat}
\bibliography{references}

\end{document}